\documentclass{jpp}
\usepackage{amsmath}
\usepackage{amssymb}
\usepackage{graphicx}
\usepackage{hyperref}
\hypersetup{hidelinks}
\usepackage{natbib}
\usepackage[utf8]{inputenc}

\newcommand{\PB}[2]{\left\{ #1 , #2 \right\}}
\renewcommand\Re{\operatorname{Re}}

\shortauthor{A. Hallenbert and G.G. Plunk}

\title{Predicting the Dimits shift through reduced mode tertiary instability analysis in a strongly driven gyrokinetic fluid limit}
\author{Axel Hallenbert\aff{1}
  \corresp{\email{axel.hallenbert@ipp.mpg.de}}
 \and Gabriel G. Plunk\aff{1}}
\affiliation{\aff{1}Max-Planck-Institut f\"ur Plasmaphysik, D-17491 Greifswald, Germany}
\date{November 2020}

\begin{document}

\maketitle

\begin{abstract}
    The tertiary instability is believed to be important for governing magnetised plasma turbulence under conditions of strong zonal flow generation, near marginal stability.  In this work, we investigate its role for a collisionless strongly driven fluid model, self-consistently derived as a limit of gyrokinetics. It is found that a region of absolute stability above the linear threshold exists, beyond which significant nonlinear transport rapidly develops. Characteristically this range exhibits a complex pattern of transient zonal evolution before a stable profile can arise. Nevertheless, the Dimits transition itself is found to coincide with a tertiary instability threshold, so long as linear effects are included. Through a simple and readily extendable procedure, tracing its origin to \citet{St-Onge2018}, the stabilising effect of the typical zonal profile can be approximated, and the accompanying reduced mode estimate is found to be in good agreement with nonlinear simulations.
\end{abstract}

\section{Introduction}

Experimental fusion devices exhibit significantly higher transport than neoclassical predictions. The additional anomalous transport arises as a result of gyroscale microturbulence driven by various instabilities \citep{Liewer1985}, such as the ion temperature gradient (ITG) mode \citep{Choi1980,Horton1981,Connor1994} or the trapped electron mode \citep{Kadomtsev1970,Nordman1990}. Moreover, the turbulent transport associated with these instabilities is very stiff. Thus, once instability is present, even a small increase in the plasma gradients will drastically increase transport levels, effectively freezing the gradients in place and restricting device performance \citep{Ryter2011}. This picture is expected to continue to hold for future fusion devices, and so being able to predict when this transport threshold is reached becomes of key importance to predict overall behaviour and performance. This is of obvious importance for the understanding and design of experiments, possibly being particularly useful for optimisation. Here, especially stellarator devices spring to mind, since they possess a large degree of freedom in their magnetic geometry \citep{Mynick2006}.

Naively, one might expect that the transport threshold should coincide with the linear instability threshold, since fundamentally these extract free energy from the plasma gradients to drive the turbulent transport. However, instead it is found that finite transport actually commences at significantly steeper gradients. This apparent discrepancy traces its origin to self-generated poloidal zonal flows \citep{Lin1998,Diamond2005}. Once the primary drift waves reach sufficient magnitude, such flows naturally arise through nonlinear interactions in what is known as a secondary instability \citep{Rogers2000b}. As the zonal flows become strong enough, they can then, in turn, nonlinearly stabilise the primary instability by shearing drift waves and decreasing their correlation length \citep{Biglari1990a}. Because the zonal flows have a Landau-undamped component \citep{Rosenbluth1998a} they can, close to marginal stability and in the absence of collisions, persist for such a long time that the effective transport nearly vanishes. This is known as the Dimits regime, and the effective upshift of the critical gradient, i.e. the difference between the linear critical gradient and the observed critical gradient for the onset of turbulence, is known as the Dimits shift, both after their discoverer \citep{Dimits2000}.

Despite the qualitative picture of the Dimits shift as just outlined being somewhat firmly established, there are still some key features which are poorly understood. Thus a general quantitative prediction of the Dimits shift has proven elusive. To describe the ITG turbulence typically observed in experiments, it is necessary to employ full gyrokinetics to retain all relevant physics \citep{Catto1978,Frieman1982,Abel2013}. This however is a highly complex kinetic system, and attempting to thoroughly account for all the possibly relevant features necessary for a full description of the Dimits shift has proved a daunting task. Instead much research has been undertaken for simpler systems which are analytically tractable, typically of the Hasegawa-Mima-Wakatani family \citep{Hasegawa1978,Hasegawa1983}, in order to gain the insight necessary to parse key features which could render the gyrokinetic problem solvable.

Many different features have been observed which could prove to be of relevance for the full problem. These include, but are not limited to, coupling to subdominant modes at unstable scales \citep{Makwana2014,Pueschel2021}, time-coherent localised soliton structures known as ferdinons \citep{VanWyk2016,VanWyk2017,Ivanov2020}, zonal-drift predator-prey-type interactions \citep{Kobayashi2012,Berionni2011}, or the ability of a turbulent momentum flux to tear down or build up a decaying zonal profile \citep{Kim2002,Ivanov2020}. One feature which however repeatedly crops up in these studies is that instability causing drift waves to arise from an initially zonally dominated state, known as the tertiary instability \citep{Rogers2000b}.

Despite seemingly being a natural candidate to explain the observed Dimits shift, based on findings from simpler systems, the importance of the tertiary instability for the Dimits shift has nevertheless been a topic of debate within the literature. \citet{St-Onge2018} and \citet{Zhu2020}, for example, based accurate predictions upon it, while \citet{Li2018a} and \citet{Ivanov2020} on the contrary reported finding it unimportant. To help rectify this confusion, in this paper we will thus attempt to shed some light on the tertiary mode in the Dimits regime, investigating its relevance for the Dimits transition in a strongly driven fluid system directly derived from gyrokinetics. 

In our investigations we will find that, just like \citet{Zhu2020} stressed, in order to properly capture the behaviour of the tertiary instability in the marginally stable regime, the linear drive cannot be neglected. The tertiary instability should not be treated as a purely shear Kelvin-Helmholtz-like (KH) instability, but instead as a modified primary instability that includes such terms. Then the tertiary instability alone seems sufficient to encapsulate the Dimits transition for the system under consideration. This is despite the fact that this system is ostensibly similar to the one recently studied by \citet{Ivanov2020}, where the opposite case was found to hold, a discrepancy arising from the present absence of collisional zonal flow damping. Finally we will see that a reduced mode scheme to approximate the tertiary instability can yield a simple but effective prediction (within 15-30\%). Furthermore this scheme seems readily extendable to more complete collisionless systems, including gyrokinetics itself, which will be the subject of an upcoming publication. 

This paper is outlined as follows. The strongly driven gyrofluid-system will first be introduced in Section \ref{BasicModel} and its key features will then be presented in \ref{KeyFeatures}. Next we will in turn describe each of the present instabilities of the primary-secondary-tertiary paradigm \citep[see][]{Kim2002}, noting their effects on the system as a whole. Guided by direct simulations presented in Section \ref{NonlinearSimulationsResults}, we will then home in further on the tertiary instability in Section \ref{ReducedModeDimitsShiftEstimate}. There we will show that it can be employed to arrive at a very simple Dimits shift estimate, related to the one of \citet{St-Onge2018}, which could prove to be broadly applicable for other non-collisional systems as well. Finally we will conclude with a brief summary and discussion in Section \ref{Discussion}.

\section{Basic model}\label{BasicModel}

The Dimits shift was originally observed in, and is of most experimental relevance for, fully gyrokinetic simulations of tokamaks \citep{Dimits2000}. However, the intrinsic kinetic nature of this system makes analytical treatment of even just the tertiary instability intractable. Investigations have therefore focused on simplified problems \citep[see e.g.][]{Kolesnikov2005a,Numata2007}, hoping to find insights which can be extrapolated to the more complete problem. Naturally these models all fail to capture much of the physics of the full gyrokinetic system because of their simplicity, possibly raising concerns about how valid such extrapolation will be. Therefore we will here present another self-consistently closed gyrofluid system in two spatial dimensions, in the hope that it may prove yet another useful stepping stone to solidify and clarify the emerging picture of the Dimits shift when proceeding towards the full gyrokinetic problem. 

\subsection{Gyrokinetics and conventions}\label{GyrokineticsAndConventions}

To arrive at the system of interest one starts from the usual electrostatic collisionless gyrokinetic equation in Fourier space, which we in the vein of \citet{Plunk2014a} express in non-dimensional form as
    \begin{equation}\label{gyrokineticequation}
        \left( \frac{\partial}{\partial t} + i \tilde{ \omega }_d \right) h_{\mathbf{k}} + \PB{\Phi}{h}_\mathbf{k} = \left( \frac{\partial}{\partial t} + i \tilde{ \omega }_* \right) \Phi_\mathbf{k} f_0.
    \end{equation}
Here $f_0$ is the ion Maxwellian distribution with mean thermal velocity $v_T = \sqrt{ 2 T / m}$, $h$ is the non-adiabatic part of the ion fluctuations $\delta f_i$. Meanwhile, the gyroaverage in Fourier space is encapsulated by the Bessel function of the first kind $J_0 = J_0 (\sqrt{2} k_\bot w_\bot)$, where the normalised velocity $w=v/v_T$ and wavenumber $k$ are split into their parallel and perpendicular components $w_\parallel, k_\parallel, w_\perp, k_\perp$ with respect to the magnetic field. It enters \eqref{gyrokineticequation} through the gyroaveraged electrostatic potential $\Phi_{\mathbf{k}}=J_0\varphi_{\mathbf{k}}$, while the Fourier space Poisson bracket, in turn, is given by
    \begin{equation}\label{PoissonBracket}
        \PB{a}{b}_\mathbf{k}=\sum_{\mathbf{k}_1,\mathbf{k}_2}\left(k_{1y}k_{2x}-k_{1x}k_{2y}\right)a_{\mathbf{k}_1}b_{\mathbf{k}_2}\delta_{\mathbf{k},\mathbf{k}_1+\mathbf{k}_2},
    \end{equation}
where $\delta_{\mathbf{k},\mathbf{k}_1+\mathbf{k}_2}$ is the Kronecker delta and the $x$- and $y$-coordinates are the radial and poloidal coordinates respectively. 
After the introduction of a reference length scale $L_{\mathrm{ref}}$, the spatial and temporal dimensions are normalised to the typical ion gyroradius $\rho$ and the streaming time $v_T / L_{\mathrm{ref}}$ respectively, so that $\varphi = q \phi L_{\mathrm{ref}}/ T \rho$ is the dimensionless electrostatic potential. Furthermore the plasma $\beta$ is assumed small so that the magnetic field $\mathbf{B}=B\mathbf{b}$ satisfies $\nabla \ln B \approx \mathbf{b} \cdot \nabla \mathbf{b}$, which enables the velocity-dependent diamagnetic and magnetic drift frequencies
    \begin{equation}
        \tilde{ \omega }_* = \omega_* \left( 1 + \eta \left( w^2-\frac{3}{2} \right) \right)\;\;\; \mathrm{and} \;\;\;\tilde{ \omega }_d = \omega_d \left( w_\parallel^2 + \frac{ w_\perp^2 }{ 2} \right)
    \end{equation}
to be succinctly expressed entirely in terms of the four parameters 
    \begin{equation}\omega_* = \frac{k_yL_{\mathrm{ref}}}{\sqrt{2}L_n} = \omega_{*0} k_y,\; \omega_d =  \frac{\sqrt{2}k_yL_{\mathrm{ref}}}{R} = \omega_{d0} k_y,\; \eta  = \frac{L_n}{L_T},\; \mathrm{and}\; \tau=\frac{T_i}{T_e},
    \end{equation}
from the electron/ion temperatures $T_{e/i}$ and the characteristic density, temperature, and magnetic curvature lengths
\begin{equation}
    L_n=\left(\frac{d\ln n}{dx}\right)^{-1},\;\;\; L_T=\left(\frac{d\ln T}{dx}\right)^{-1},\;\;\; \mathrm{and} \;\;\; R=\left(\frac{d\ln B}{dx}\right)^{-1},
\end{equation}
all of which are negative by our convention.

To couple the potential $\varphi$ to the ion gyrocentre distribution $h$ and close the system, the electrons are taken to follow a modified adiabatic response \citep{Dorland1993,Hammett1993} such that the quasineutrality condition becomes
    \begin{equation}\label{quasineutrality}
        \int d^3\mathbf{v} J_0 h_\mathbf{k} = n (1 + \tau \hat{\alpha}) \varphi_\mathbf{k},
    \end{equation}
where $\hat{\alpha}$ is the operator
    \begin{equation}\label{alphaoperator}
        \hat{\alpha}a=a(x,y)-\frac{1}{L_y}\int_0^{L_y}a(x,y)dy,
    \end{equation}
i.e. an operator that is zero when acting on purely zonal $\mathbf{E}\times \mathbf{B}$ modes with $k_y=0$, and unity otherwise.

To serve our purpose of studying the Dimits shift, the gyrokinetic equation in the form of \eqref{gyrokineticequation} clearly neglects both parallel variations and collisions. The former omission constitutes a considerable simplification from a spatially 3D to a spatially 2D system, but necessarily excludes the ITG slab mode. Instead the focus becomes a local description of the well-known bad-curvature-driven toroidal ITG instability \citep{Beer1995a}, which seems to be of most relevance for the Dimits transition \citep{Dimits2000}. The second omission is similarly made because, should collisions be included, their presence significantly muddles the waters. This is because a wide range of zonal flow behaviour then manifests, including bursty patterns \citep[see][]{Berionni2011} or non-quasistatic flows \citep[see][]{Kobayashi2012}, so that it can become somewhat difficult to identify a clear Dimits transition or even reliably define the Dimits shift. However, in their absence, Landau-undamped Rosenbluth-Hinton states \citep{Rosenbluth1998a} can produce static zonal flow states with zero transport, in principle (only limited by the finite simulation time available to find such a state) providing a clear cut distinction between systems within and outside the Dimits regime.

\subsection{The strongly driven gyrokinetic fluid limit}\label{StronglyDrivenGyrokineticFluidLimit}

Employing a subsidiary ordering such that 
    \begin{equation}\label{SubsidiaryOrdering}
        \frac{\partial_t}{\eta\omega_*} \sim \frac{\omega_d}{\omega_*} \sim k_\bot^2 \sim  \frac{\tilde{\varphi}^2}{\overline{\varphi}^2} \sim \frac{1}{\eta} \ll 1,
    \end{equation}
where the gyrophase-independent response and potential have been split into their zonal and nonzonal components like
    \begin{equation}
        \overline{h} = \left(1-\hat{\alpha}\right) h, \;\; \;\;\tilde{h} = \hat{\alpha} h, \;\;\; \overline{\varphi} = \left(1-\hat{\alpha}\right) \varphi, \;\; \;\;\tilde{\varphi} = \hat{\alpha} \varphi,
    \end{equation}
one finds (see Appendix \ref{Derivation}) that the gyrokinetic moment hierarchy self-consistently closes at second order, resulting in the renormalised equation system:
    \begin{align}\label{phi_drift}
        &\frac{\partial \tau \tilde{\varphi}_{\mathbf{k}}}{\partial t} + \PB{\overline{\varphi}}{\tau \tilde{\varphi}}_\mathbf{k} + \PB{\nabla_\bot^2 \overline{\varphi}}{\tilde{T}_\perp}_\mathbf{k} - k_\bot^2 \PB{\overline{\varphi}}{\tilde{T}_\perp}_\mathbf{k} -
        \PB{\overline{\varphi}}{\nabla_\bot^2\tilde{T}_\bot}_\mathbf{k}\nonumber \\
        &= i \omega_* \left(1  - \eta k_\bot^2 \right) \tilde{\varphi}_{\mathbf{k}} - i \omega_d \left( \frac{\tilde{T}_{\parallel\mathbf{k}}}{2} + \frac{\tilde{T}_{\perp\mathbf{k}}}{4} \right) - D_\mathbf{k} \tau \tilde{\varphi}_\mathbf{k},
    \end{align}
    \begin{equation}\label{phi_zonal}
        \frac{\partial k_x^2 \overline{\varphi}_\mathbf{k}}{\partial t} + \PB{\nabla_\bot^2 \tilde{\varphi}}{\tilde{T}_\perp}_\mathbf{k} - k_x^2 \PB{\tilde{\varphi}}{\tilde{T}_\perp}_\mathbf{k} - \PB{\tilde{\varphi}}{\nabla_\bot^2 \tilde{T}_\bot}_\mathbf{k} = 0,
    \end{equation}
    \begin{equation}\label{t_perp}
        \frac{\partial \tilde{T}_{\perp\mathbf{k}}}{\partial t} + \PB{\overline{\varphi}}{\tilde{T}_\perp}_\mathbf{k} = \frac{i \omega_* \eta \tilde{\varphi}_\mathbf{k}}{2} - D_\mathbf{k} \tilde{T}_{\perp\mathbf{k}},
    \end{equation}
    \begin{equation}\label{t_par}
        \frac{\partial \tilde{T}_{\parallel\mathbf{k}}}{\partial t} + \PB{\overline{\varphi}}{\tilde{T}_\parallel}_\mathbf{k} = \frac{i \omega_* \eta \tilde{\varphi}_\mathbf{k}}{4} - D_\mathbf{k} \tilde{T}_{\parallel\mathbf{k}}.
    \end{equation}
Here an ad hoc damping operator $D_\mathbf{k}$ acting on the nonzonal components, to be further discussed in Section \ref{TheDampingOperator}, has been added. This is to compensate for the loss of collisionless damping \citep{Landau1946} that occurs upon taking moments of the gyrokinetic equation. Note that the zonal components of the temperature do not enter, the system only consists of one zonal field $\bar{\varphi}$ and three nonzonal fields $\tilde{\varphi}$, $\tilde{T}_\perp$, $\tilde{T}_\parallel$, which, as a consequence of \eqref{SubsidiaryOrdering}, differ by order like
\begin{equation}
    \frac{\tilde{\varphi}}{\tilde{T}_\perp} \sim \frac{\tilde{\varphi}}{\tilde{T}_\parallel} \sim \frac{1}{\eta}.
\end{equation} 
However, combining \eqref{t_perp} and \eqref{t_par} it is clear that the volume average of $\delta T=\tilde{T}_{\perp}-2\tilde{T}_{\parallel}$ transiently decays to zero under the action of $D_\mathbf{k}$. Nevertheless, we include this component in our simulations for completeness.

Some comments about \eqref{SubsidiaryOrdering} and its resulting system are now in order. First, apart from the additional separation of zonal and nonzonal components in the ordering scheme, this corresponds to a strongly driven limit with a high temperature gradient feeding a strong ITG instability and causing long wavelength turbulence to be dominant, previously studied separately in its linear \citep{Plunk2014a} and nonlinear \citep{Plunk2012} limits. Note that though we call the limit ``strongly driven" since the drive term is large compared to the particle drift, stable modes still do exist, so one might alternatively call this limit nonresonant (in the linear fluid sense). As to the specific additional zonal/nonzonal separation within the ordering scheme, it is necessary for a consistent closure which includes both linear and nonlinear interactions. Beyond this it also encapsulates the fact that only the former are so called modes of minimal inertia \citep{Diamond2005}, being easily excited due to the density shielding of the adiabatic electron response. Furthermore, being Landau-undamped they can persist for long times, and so they are observed to be comparatively strong.

Secondly, all the present nonlinear terms affecting the drift waves involve zonal flows. \citet{Farrell2009} have already shown that, beyond the Dimits regime, simple systems (specifically Hasegawa-Wakatani) can exhibit all relevant physics despite lacking drift wave self-interactions. Thus it may be unsurprising that we here will find that the same can be true inside the Dimits regime. Beyond this we note that the full nonlinear interaction is asymmetrical between the different fields. While the governing equations for the nonzonal fields all include the typical $\mathbf{E}\times \mathbf{B}$-advection nonlinear $\PB{\overline{\varphi}}{\cdot}$-term, both the zonal and nonzonal potentials $\overline{\varphi}$ and $\tilde{\varphi}$ are affected by an additional set of nonlinear diamagnetic drift FLR terms coupling them to $\tilde{T}_\bot$. It should also be noted that by ordering there is no  Reynolds stress, i.e. a term of the form $\PB{\varphi}{\nabla^2\varphi}_\mathbf{k}$, present. It has been pointed out that such a term greatly facilitates the construction of strong zonal flows \citep{Diamond1991}, but as a consequence of zonal flows being unaffected by $D_\mathbf{k}$, zonally dominated states will here arise even though the Reynolds stress is absent.

Thirdly, barring the splitting of the temperature moment into its separate parallel and perpendicular components, the nonlinear interaction is the same as \citet{Plunk2012}. By a trivial modification of the results therein, the electrostatic energy conserved by the nonlinear interactions in \ref{phi_drift}-\ref{t_par} is therefore readily found to be given by
\begin{equation}\label{EnergyDefinition}
    E_\varphi = E_{\overline{\varphi}} + E_{\tilde{\varphi}} = \frac{1}{2L_xL_y}\int |\nabla\overline{\varphi}|^2dxdy + \frac{1}{2L_xL_y}\int \tau\tilde{\varphi}^2 dxdy.
\end{equation}

Finally, this strongly driven system seems formally far from the usual marginally unstable Dimits regime by virtue of its ordering, and one might question its relevance when investigating the Dimits shift. However, with a sufficiently large $D_\mathbf{k}$, marginal stability can be reinstated and a clear Dimits regime emerges. This system may thus act as a stepping stone, since its self-consistent closure means that the nonlinear interaction should closely resemble that of full gyrokinetics, at least in its range of validity. Indeed it bears much resemblance to another, but highly collisional, gyrokinetic fluid limit recently studied by \citet{Ivanov2020}. Beyond being collisionless, it differs in mainly three ways: i) zonal flows are not subject to collisional dissipation  ii) the nonlinear drift wave self-interaction and Reynolds stress become too small to be of relevance, iii) the zonal temperature perturbations cease to be dynamically relevant.

\section{Key features}\label{KeyFeatures}

\subsection{Primary instability}\label{PrimaryInstability}

To arrive at a linear dispersion relation for the primary modes of the system \eqref{phi_drift}-\eqref{t_par}, plane wave solutions proportional to $\exp\left(\lambda^p_\mathbf{k}t + i \mathbf{k}\cdot\mathbf{r}\right)$ are postulated, where $\lambda^p_\mathbf{k}$ can be split into the growth rate $\gamma^p_\mathbf{k}$ and frequency $\omega^p_\mathbf{k}$ according to $\lambda^p_\mathbf{k} = \gamma^p_\mathbf{k} - i \omega^p_\mathbf{k}$. A straightforward linear instability calculation then reveals the presence of a pure temperature mode (where $\tilde{\varphi}_\mathbf{k}=0$, but $\tilde{T}_\perp$ and $\tilde{T}_\parallel$ are nonzero) which is strictly damped, 
\begin{equation}\label{DampedMode}
    \gamma^{pT}_\mathbf{k}=-D_\mathbf{k},
\end{equation}
and two modes with the expected dispersion forms 
    \begin{equation}\label{LinearGrowth}
        \lambda_\mathbf{k}^{p\pm} = \gamma_\mathbf{k}^{p\pm} - i \omega_\mathbf{k}^{p\pm} = - D_\mathbf{k} + i \omega_* \frac{1 - \eta k_\bot^2}{2 \tau} \pm \frac{1}{2 \tau} \sqrt{ \eta \tau \omega_* \omega_d - \omega_*^2 \left( 1 - \eta k_\bot^2 \right)^2 }
    \end{equation}
of the toroidal ITG mode inherent to the ordering \eqref{SubsidiaryOrdering} \citep{Plunk2014a}. Note that here and elsewhere we reserve the $p$, $s$, $t$ superscripts for primary, secondary, and tertiary quantities, and use $\pm$-superscripts to indicate the most/least unstable modes of each kind. 

\begin{figure}
\centering
\includegraphics[width=0.9\linewidth]{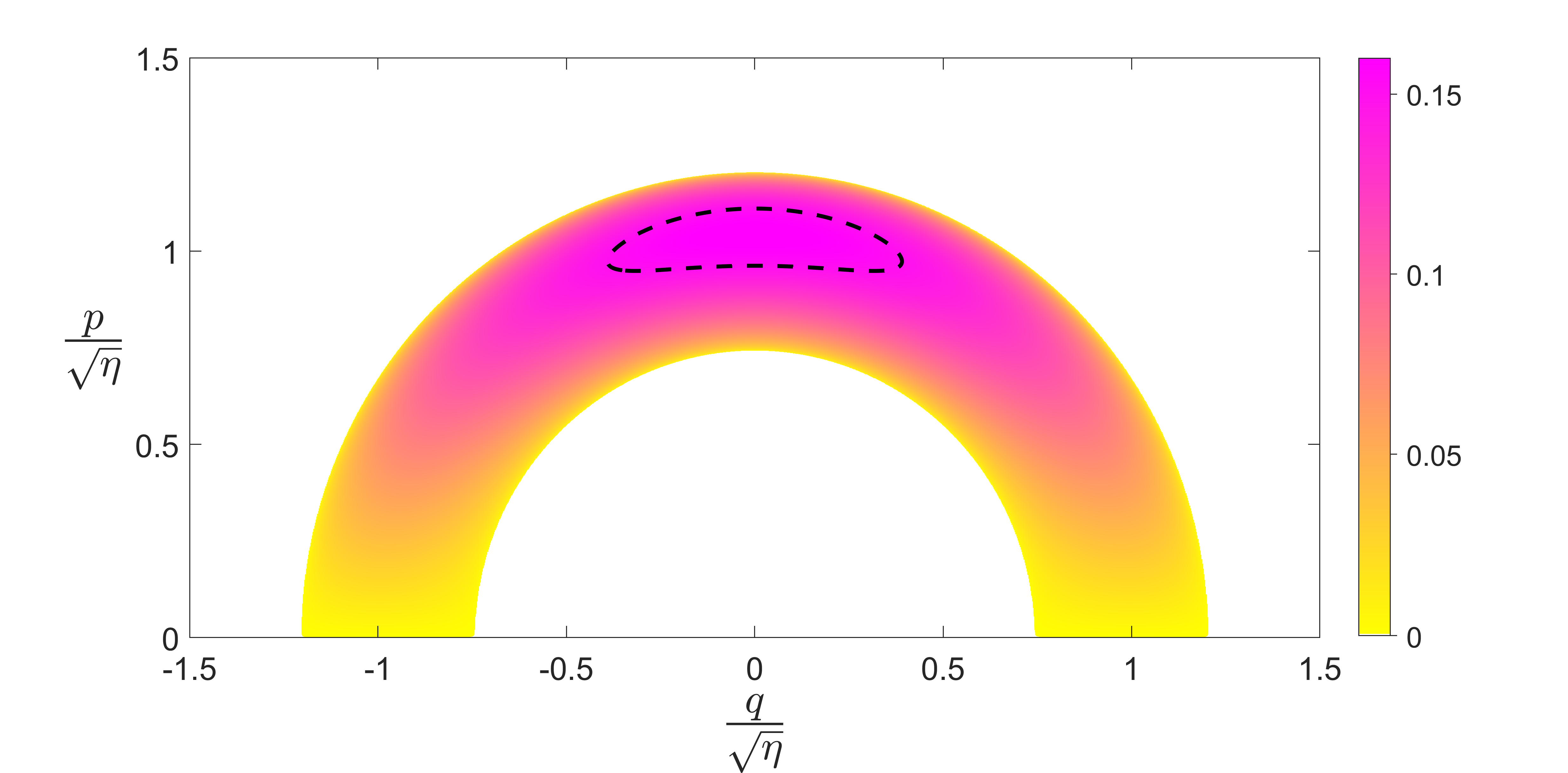}
\caption{Normalised primary drift wave growth rate $-\omega_{*0}^{-1}\gamma_\mathbf{k}^{p+}$ from \eqref{LinearGrowth} as a function of the radial and poloidal wavenumbers $q$ and $p$ for $\eta\tau\omega_d/\omega_*=0.2$, showing clearly the instability annulus \eqref{InstabilityAnnulus}. The instability boundary for primary unstable modes, in the presence of that $D_\mathbf{k}=D$ for which this configuration constitutes the Dimits threshold, is also shown.}
\label{LinearGrowthFigure}
\end{figure}

Because $D_\mathbf{k}$ generally introduces only a $k$-dependent shift in $\gamma$ towards lower values, it is useful to first consider $D_\mathbf{k} = 0$. Remembering that the definitions of $\omega_*$ and $\omega_d$ include a factor $k_y$, several features are readily apparent. The most unstable mode is as expected the purely radial streamer with $k_x=q=0$ satisfying
    \begin{equation}\label{MaxUnstableky}
        k_y^2 = p^2 = \frac{1}{3\eta}\left(2+\sqrt{1 + \frac{3 \eta \tau \omega_d}{\omega_*}}\right).
    \end{equation}
Note here the introduction of $q$ and $p$ which will henceforth be used for poloidal and radial wavenumbers respectively. Now, when \eqref{MaxUnstableky} is inserted into \eqref{LinearGrowth} it gives the expected bad curvature ITG instability scaling \citep[see][]{Beer1995a}
    \begin{equation}
        \gamma_\mathbf{k}^{p+} \propto \frac{k_y}{\sqrt{\tau R L_T}},
    \end{equation}
when the correction term under the root is taken to be small. When this term on the other hand is sufficiently large, the growth rate passes through zero. Thus we find that only the wavenumbers within the annulus 
    \begin{equation}\label{InstabilityAnnulus}
        \frac{1}{\eta}\left(1 - \sqrt{\frac{\eta \tau \omega_d}{\omega_*}}\right) < k_\bot^2 < \frac{1}{\eta}\left(1 + \sqrt{\frac{\eta \tau \omega_d}{\omega_*}}\right)
    \end{equation}
can be unstable. Here we see that $\eta$ pushes the instability to larger scales, while $\tau\omega_d/\omega_*$ controls the narrowness of the instability annulus and whether large scales are damped or not (indeed as $\eta\tau\omega_d/\omega_*$ exceeds 1, the annulus becomes a disk), see Figure \ref{LinearGrowthFigure}. Clearly \eqref{MaxUnstableky} and \eqref{InstabilityAnnulus} sets the energy injection scale to be $\sqrt{(1/\eta)}$ in accordance with the subsidiary ordering \ref{SubsidiaryOrdering}, which is therefore justified a posteriori.

We can arrive at a linear instability threshold for the temperature gradient when $D_\mathbf{k}=D$ is constant. Remembering that $\omega_*$ and $\omega_d$ both include a factor $k_y$, we find upon inserting \eqref{MaxUnstableky} into \eqref{LinearGrowth} and setting the result to $0$, that it becomes a condition on $\eta(D)$ for the most unstable mode to be marginally stable. Its $\eta>0$-solution is denoted by
    \begin{equation}\label{SystemThreshold}
        \eta^0 = \frac{8 \tau D^2}{9\omega_{d0}\omega_{*0}}
    \end{equation}
and it is found that, for $\eta$ larger than this value, $\gamma^{p+}$ increases monotonically with $\eta$. When $D_\mathbf{k}$ is allowed to vary with respect to $p$, a correction to \eqref{SystemThreshold} appears, but monotonicity continues to hold. Therefore, for some $\eta^0$ (typically close to \eqref{SystemThreshold} with $D=D_\mathbf{p}$) we have the necessary condition for instability
    \begin{equation}
        \eta > \eta^0.
    \end{equation}

Returning again to the unstable mode it naturally includes both potential and temperature perturbations, so upon inserting \eqref{LinearGrowth} into \eqref{phi_drift}-\eqref{t_par} the ratio between the two can be calculated to be
    \begin{equation}\label{ModePhase}
        \frac{\tilde{\varphi}_\mathbf{k}}{\tilde{T}_{\perp\mathbf{k}}} = \frac{\tilde{\varphi}_\mathbf{k}}{2\tilde{T}_{\parallel\mathbf{k}}} = \frac{1}{\eta\tau}\left(1-\eta k_\bot^2 - \sqrt{(1-\eta k_\bot^2)^2 - \frac{\eta\tau\omega_d}{\omega_*}}\right).
    \end{equation}
Using \eqref{InstabilityAnnulus} to parametrise all the unstable modes with an angle $0\leq\theta\leq\pi$ like
    \begin{equation}
        k_\bot^2 = \frac{1}{\eta}\left(1 - \cos\theta\sqrt{\frac{\eta \tau \omega_d}{\omega_*}}\right),
    \end{equation}
one finds upon inserting this into the RHS of \eqref{ModePhase} that it reduces to the simple expression
    \begin{equation}\label{NiceModeExpression}
        \frac{\tilde{\varphi}_\mathbf{k}}{\tilde{T}_{\perp \mathbf{k}}} = \frac{\tilde{\varphi}_\mathbf{k}}{2\tilde{T}_{\parallel \mathbf{k}}} = \sqrt{\frac{\omega_d}{\eta\tau\omega_*}}e^{ik_y\theta/|k_y|}
    \end{equation}
since by our convention $\omega_{*0}<0$. Now since the radial heat flux is given by
    \begin{equation}\label{HeatFlux}
        Q=\frac{1}{L_xL_y}\int \hat{x}\cdot v_{\mathbf{E}} (\tilde{T}_\bot + \tilde{T}_\parallel) dy dx = \sum_{\mathbf{k}}\Re{\left(ik_y\tilde{\varphi}^*_\mathbf{k}(\tilde{T}_{\perp \mathbf{k}}+\tilde{T}_{\parallel \mathbf{k}})\right)},
    \end{equation}
\eqref{NiceModeExpression} implies that each mode provides a positive contribution to the total heat flux, since
    \begin{equation}
        \Re{\left(ik_y\tilde{\varphi}^*_\mathbf{k}(\tilde{T}_{\perp \mathbf{k}}+\tilde{T}_{\parallel \mathbf{k}})\right)}=3\sqrt{\frac{\omega_d}{\eta\tau\omega_*}}|\tilde{\varphi}_\mathbf{k}|^2k_y\sin{\left(\frac{k_y\theta}{|k_y|}\right)} \geq 0.
    \end{equation}
Additionally, \eqref{NiceModeExpression} makes it clear that the potential component of the mode decreases compared to its temperature as $\eta$ increases. This linear result can be approached intuitively, since a strong temperature gradient causes $v_\mathbf{E}\cdot\nabla f_0$, the free energy source term in the gyrokinetic equation (where $v_\mathbf{E}$ is the $\mathbf{E}\times \mathbf{B}$-drift), to possess a larger temperature moment than density moment, the latter of which is most important for $\varphi$ through the quasineutrality condition \eqref{quasineutrality}.

\subsection{The damping operator}\label{TheDampingOperator}

Having determined the linear properties of Equations \eqref{phi_drift}-\eqref{t_par} we are now in a position to discuss $D_\mathbf{k}$ in greater detail. Examining the linear growth rate \eqref{LinearGrowth} it is clear that, as long as there exists bad magnetic curvature providing finite $\omega_d$, and in the absence of artificial dissipation $D_\mathbf{k}$, the primary instability is present at $\eta k_\bot^2=1$ given any arbitrarily small density and temperature gradients. Furthermore all arbitrarily small scales are completely undamped and so can act as a reservoir of energy. Numerically, this means that even though every unstable $\lambda_\mathbf{k}^{p+}$-mode in the injection range is accompanied by a damped $\lambda_\mathbf{k}^{p-}$-mode, without $D_\mathbf{k}$ the system could nonlinearly diverge while exhibiting large-scale energy pileup typical of 2D-turbulence with its inverse energy cascade \citep{Kraichnan1967,Qian1986,Terry2004}. In 3D turbulence this is prevented by a scale balance of parallel streaming and turbulence known as critical balance \citep{Barnes2011}, but no such mechanism is available here.

Given what was just outlined, in order to prevent nonphysical absolute instability and the excitation of arbitrarily small or large scales, the necessity of including some kind of $D_\mathbf{k}$ is apparent. Physically this is meant to represent the Landau-type damping present in weakly collisional toroidal ITG but which was lost upon only considering its moments to arrive at \eqref{phi_drift}-\eqref{t_par} \citep{Sugama1999}. Though our ordering \eqref{SubsidiaryOrdering} implies that the kinetic damping is small, it is nevertheless non-zero and so dynamically relevant, particularly for the marginally stable small scales it firmly stabilises. Its inclusion is further justified since the ordering \eqref{SubsidiaryOrdering}, though ``strongly driven", nevertheless allows the primary instability to also be weak, so that $D_\mathbf{k}$ can stabilise the system so it exhibits a Dimits regime. 

As to the specific form of $D_\mathbf{k}$ which we will employ in this paper, we will always include a constant component $D$, present for all nonzonal modes. The reason why is that, at least within and close beyond the Dimits regime, we have found its inclusion to be sufficient to prevent a large-scale energy pileup. This form has some physical justification, in that a rigorous linear analysis of the full kinetic mode reveals, beyond the normal mode whose Landau damping can be approximated as viscous dissipation, the presence of an algebraically decaying continuum mode. In the marginally stable regime of interest the continuum modes of the sidebands should thus be dominant, since they decay much more slowly \citep{Sugama1999,Mishchenko2018}. Unfortunately it is very hard to accurately reproduce the behaviour of these modes in a non-exotic way for a spectral fluid model \citep{Sugama1999}, and so in absence of better alternatives a flat decay can be used to model this. Beyond this component it is natural to include some kind of hyperviscosity $k_\bot^\alpha$ in $D_\mathbf{k}$. However, this seems to have little effect on the key results of this paper, presumably because of how sharply peaked in $k$-space the linear instability is, and so we typically do not include it. For generality we will nevertheless allow it in our instability calculations.

\subsection{Secondary instability}\label{SecondaryInstability}

Within the primary-secondary-tertiary hierarchy, the secondary instability develops once the primary drift waves have grown sufficiently for slight flow inhomogeneities to amplify through a shearing interaction to magnify small zonal perturbations \citep{Kim2002}. It is analytically best treated via a Galerkin truncation of \eqref{phi_drift}-\eqref{t_par} into the 4-mode system consisting of the most unstable purely radial primary mode, its two sidebands, and a single zonal mode: 
    \begin{equation}\label{4MGalerkin}
        \mathbf{p}=(0,p),\;\mathbf{r}^\pm=(\pm q,p),\;\mathbf{q}=(q,0).
    \end{equation}
The potential and temperature amplitudes of the primary mode are then fixed and taken to be much larger than other variables so that the linear terms of the sidebands can be ignored compared to the nonlinear interaction with the primary mode, which now become linearised. Then it is straightforward to obtain the KH-like secondary dispersion relation
    \begin{equation} \label{SecondaryAlmostDispersion}
        {\lambda^{s}}^2 + 4 p^2 q^2 \left| \tilde{T}_{\bot \mathbf{p}} \right|^2 \left( \frac{2 q^2 }{\tau} - \mathrm{Re}\left(\frac{\tilde{\varphi}_\mathbf{p}}{\tilde{T}_{\bot \mathbf{p}}}\right)\right) = 0.
    \end{equation}
    
Inserting the potential/temperature ratio of the unstable mode \eqref{ModePhase} into \eqref{SecondaryAlmostDispersion} is now natural since it is this mode which initiates the secondary instability in the primary-secondary paradigm, and this results in
    \begin{equation}\label{SecondaryDispersion}
        {\lambda^s}^{2} + \frac{4 p^2 q^2 \left| \tilde{T}_{\bot \mathbf{p}} \right|^2}{\eta\tau} \left( 2\eta q^2 + \eta p^2 - 1 + \mathrm{Re}\sqrt{(1-\eta p^2)^2-\frac{\eta\tau\omega_d}{\omega_*}} \right) = 0.
    \end{equation}

\begin{figure}
\centering
\includegraphics[width=0.75\linewidth]{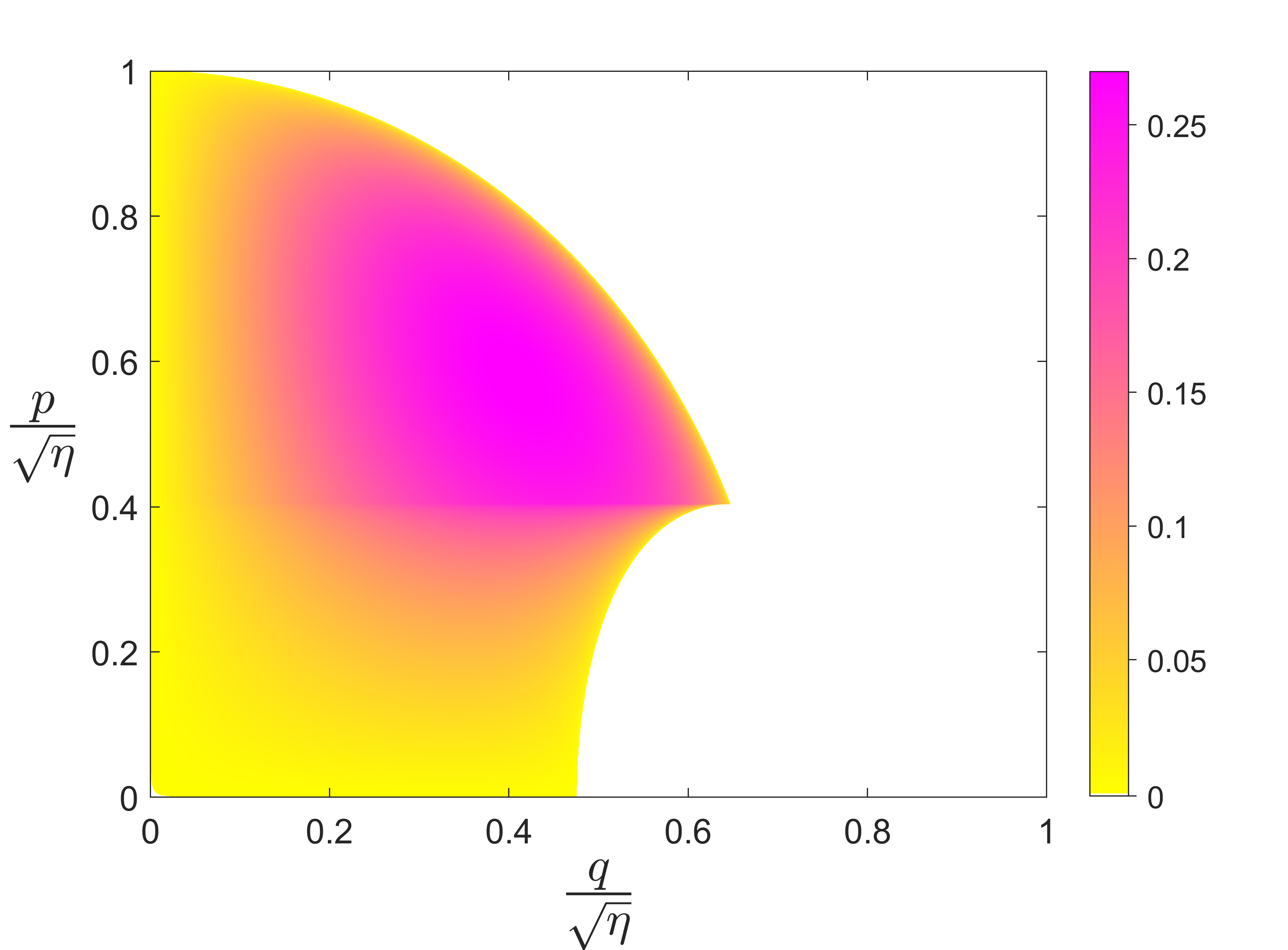}
\caption{Normalised secondary growth rate $\sqrt{\tau\eta}|\tilde{T}_{\perp\mathbf{p}}|^{-1}\gamma^{s+}$ as a function of $q$ and $p$ for $\eta\tau\omega_d/\omega_*=0.7$. The bifurcation of \eqref{SecondaryBifurcation} is clearly visible.}
\label{SecondaryInstabilityFigure}
\end{figure}

The last, stabilising term in \eqref{SecondaryDispersion} is similar to the opposite of the destabilising term of $\gamma_\mathbf{k}^{p+}$ in \ref{LinearGrowth}, but gives rise to a discontinuity instead of a bifurcation at
\begin{equation}\label{SecondaryBifurcation}
    \eta p^2 = 1\pm\sqrt{\frac{\eta\tau\omega_d}{\omega_*}},
\end{equation}
a feature which can clearly be seen in Figure \ref{SecondaryInstabilityFigure} where the secondary growth rate $\gamma^{s+}$ is plotted. As $p$ approaches the lower threshold from below, the radical in \eqref{SecondaryDispersion} approaches $0$, causing $\gamma^{s+}$ to rapidly increase until its derivative with respect to $p$ discontinuously flattens.   

Now the absolute requirement on the zonal wavenumber in order for an unstable primary mode to be secondary unstable is established by equation \eqref{SecondaryDispersion} to be
    \begin{equation} \label{SecondaryCondition}
        q^2 < \frac{1-\eta p^2}{2\eta}.
    \end{equation}
This means that modes with $p^2>\eta$, and in particular the most unstable mode satisfying \eqref{MaxUnstableky}, are completely stable to the secondary instability and so can continue to grow unabated until another channel for zonal flow generation is established. 

One might suspect that the zonal flow would be initiated by those less unstable primary modes with smaller $p$ which are able to satisfy \eqref{SecondaryCondition} once they have grown to a sufficient amplitude. However, in simulations it is instead observed that, since the primary growth rate is rather sharply peaked around \eqref{MaxUnstableky}, these modes do not grow fast enough to be dynamically relevant at this stage. Instead, it seems that, since the small $q$-sidebands of the most unstable primary mode grow at nearly the same rate, it is their mutual sideband-sideband-interaction which jump-starts the zonal growth. This is evidenced by the fact that the initial zonal growth rate remains mostly unchanged even when all modes but the most unstable primary mode and its sidebands are set to 0. Thus we conclude that the secondary instability of this form is, in fact, presently irrelevant in the Dimits regime.

\subsection{Local Tertiary instability}\label{TertiaryInstabilityLocal}

Turning now to the final stage of the primary-secondary-tertiary hierarchy, once the zonal flow has grown enough to quench the drift waves the tertiary instability is that instability which allows the drift waves to reemerge from a zonally dominated state \citep{Rogers2000b}. In analysing this instability we will consider two separate limits, one localised and one de-localised (i.e. localised in $k$-space). 

It is a well-known feature of tertiary modes that they localise to regions of zero zonal shear rate, $\partial_x^2\overline\varphi = 0$ \citep{Kobayashi2012,Kim2018,Kim2019}. Therefore we consider a poloidal band of modes ($k_y=p$) subject to a large amplitude zonal flow localised around such a point. Taking $D_\mathbf{k}=D$ again, in real $x$-space equations \eqref{phi_drift}-\eqref{t_par} then become
\begin{equation}
    \left(\partial_t+ip\partial_x\overline{\varphi}\right)\tilde{T}_\bot - \frac{i\eta\omega_*\tilde{\varphi}}{2} = -D\tilde{T}_\perp,
\end{equation}
\begin{equation}
    \left(\partial_t+ip\partial_x\overline{\varphi}\right)\tilde{T}_\parallel - \frac{i\eta\omega_*\tilde{\varphi}}{4} = -D\tilde{T}_\parallel,
\end{equation}
\begin{align}
    &\tau\left(\partial_t+ip\partial_x\overline{\varphi}\right)\tilde{\varphi} - i\omega_*\left(1-\eta(p^2-\partial_x^2)\right)\tilde{\varphi}\nonumber\\ &+i\left(\frac{\omega_d}{4}+2p\partial_x^3\overline{\varphi}+2p\partial_x^2\overline{\varphi}\partial_x\right)\tilde{T}_\bot + i\frac{\omega_d}{2}\tilde{T}_\parallel = -D\tilde{\varphi}.
\end{align}
If we consider a narrow region in which $\partial_x\overline{\varphi}$ and $\partial_x^3\overline{\varphi}$ are approximately constant and allow ourselves to consider the mode to also be localised around the most unstable primary mode with $k_x\approx 0$, we therefore find the local tertiary dispersion form
\begin{equation}\label{TertiaryDispersionLocal}
    \lambda^{t\pm}_L = -D+i\left(\frac{\omega_*-\eta\omega_*p^2}{2\tau}-p\partial_x\overline{\varphi}\right) \mp \frac{\omega_*}{2\tau}\sqrt{\frac{\eta\tau}{\omega_*}(\omega_d+8p\partial_x^3\overline{\varphi})-\left(1-\eta p^2\right)^2}.
\end{equation}

As is easily seen, this expression is precisely the linear dispersion of the primary mode \eqref{LinearGrowth} Doppler shifted by $p\partial_x\overline{\varphi}$ and with a zonal shear modified magnetic curvature. We note that the real part of this expression vanishes when the driving gradients are removed, meaning that no tertiary instability exists at all in their absence. We are thus dealing here with only a modified primary, extracting energy from the background gradients, rather than from the zonal flow like the KH instability \citep[see][]{Zhu2020}. This is because the fundamental ordering \eqref{SubsidiaryOrdering} eliminates both the Reynolds stress and the zonal temperature. If present, the former would give rise to a true tertiary KH instability \citep{Kim2002,Zhu2018}, and the latter a tertiary KH-like instability, analogous to the secondary instability \eqref{SecondaryDispersion} \citep{Rogers2000b,Ivanov2020}.

Returning to the specific expression \eqref{TertiaryDispersionLocal}, we see that the tertiary instability is asymmetric with respect to zonal flow velocity minima, $\partial^3_x\overline{\varphi}>0$, and maxima, $\partial^3_x\overline{\varphi}<0$; the former is destabilising while the latter is stabilising. This asymmetry matches gyrokinetic observations that zonal flow minima are significantly more prone to turbulent transport \citep{McMillan2011}, and has already been noted in previous tertiary instability studies of simple systems \citep[see e.g.][]{Zhu2020}. As we will see in Section \ref{NonlinearSimulationsResults}, the same holds true here: turbulence consistently localise around the points where $\partial_x^2\overline{\varphi}=0$ and $\partial_x^3\overline{\varphi}>0$ in accordance with \eqref{TertiaryDispersionLocal}. Nevertheless, results like \eqref{TertiaryDispersionLocal} can not be taken at face value. In the closely related system of \citet{Ivanov2020}, where the presence of zonal temperature perturbations considerably complicates the picture, the equivalent expression for the growth rate fails to match what is observed in simulations.

\subsection{4-Mode Tertiary Instability}\label{TertiaryInstability4M}

We now proceed to study the tertiary instability of a sinusoidal profile corresponding to the mode $\overline{\varphi}_\mathbf{q}$. In order to gain further insight we employ, for simplification, the same 4-mode (4M) Galerkin truncation \eqref{4MGalerkin} as we did for the secondary instability, even though it in general it is less justifiable here. Naturally with so few modes this analysis will fail to capture any intricate localisation effects, but we will nevertheless be able to discern some important features of the tertiary instability. After all, \eqref{TertiaryDispersionLocal} employed strong approximations (e.g. fully neglecting $k$-space coupling), which may in general not be satisfied.

Assuming without loss of generality that $\overline{\varphi}_\mathbf{q}$ is real (all results herein will only depend upon its magnitude), i.e.
\begin{equation}\label{4MZonalProfile}
    \overline{\varphi}=2\overline{\varphi}_\mathbf{q}\cos{qx},
\end{equation}
the 4M tertiary dispersion relation can after some algebra, analogous to that of Section \ref{SecondaryInstability}, be expressed as the product of three polynomials like
\begin{equation}\label{TotalTertiaryDispersion}
    \mathcal{D}_{\mathrm{P}\mathbf{r}}\left(\lambda^t_{4M}\right)\mathcal{D}_{T}\left(\lambda^t_{4M}\right)\mathcal{D}_{\mathrm{MP}}\left(\lambda^t_{4M}\right)=0.
\end{equation}
This factorised form of the dispersion relation separates the linear tertiary modes into different groups, corresponding to zeros of each of the three factors. The equation $\mathcal{D}_{\mathrm{P}_\mathbf{r}}=0$ is the unmodified primary dispersion relation of the sidebands $\mathbf{r}^\pm$ with solutions \eqref{DampedMode} and \eqref{LinearGrowth}, corresponding to a solution of \eqref{phi_drift}-\eqref{t_par} where the primary mode is absent (i.e. the $\mathbf{p}$-mode is 0) and the sidebands are of equal amplitude. Next
\begin{equation}\label{TemperatureKH}
    \mathcal{D}_{T}=(\lambda^t_{4M}+D_\mathbf{p})(\lambda^t_{4M}+D_\mathbf{r})+2p^2q^2\overline{\varphi}_\mathbf{q}^2=0
\end{equation}
is the dispersion relation of two stable pure temperature modes affected by the zonal flow, and finally
\begin{equation}\label{TertiaryDispersion4M}
    \mathcal{D}_{\mathrm{MP}} = (\lambda^t_{4M}-\lambda^{p+}_\mathbf{p})(\lambda^t_{4M}-\lambda^{p-}_\mathbf{p})(\lambda^t_{4M}-\lambda^{p+}_\mathbf{r})(\lambda^t_{4M}-\lambda^{p-}_\mathbf{r})+4p^4q^4\overline{\varphi}_\mathbf{q}^4+Ap^2q^2\overline{\varphi}_\mathbf{q}^2=0,
\end{equation}
where
\begin{align}
    A(\lambda^t_{4M}) =& \;4(\lambda^t_{4M}+D_\mathbf{p})(\lambda^t_{4M}+D_\mathbf{r})
    +\frac{\eta\omega_*\omega_d}{\tau}-2\frac{\omega_*^2}{\tau^2}(1-\eta p^2)^2\nonumber\\&
    +2i\frac{\omega_*}{\tau}\left[(\eta p^2-1)(2\lambda^t_{4M}+D_\mathbf{p}+D_\mathbf{r})-\eta q^2(\lambda^t_{4M}+D_\mathbf{r})\right],
\end{align}
is the dispersion relation of the 4M zonal flow modified primary mode. Henceforth we focus on this latter equation, since the modified primary will prove to be the most unstable tertiary mode.

Let us consider the dispersion relation of the modified primary in the large zonal flow limit. Expanding \eqref{TertiaryDispersion4M} in orders of $\overline{\varphi}_\mathbf{q}$ like
\begin{equation}
    \lambda^t_{4M}=\lambda_1^t+\lambda_{1/2}^t+\textit{O}(1),
\end{equation}
we find, after collecting terms up to to order $\sqrt{\overline{\varphi}_\mathbf{q}}$ and using \eqref{LinearGrowth}, that \eqref{TertiaryDispersion4M} can be reduced to
\begin{equation}\label{RootPhiExpansion}
    \left(\left(\lambda^t_{4M}\right)^2+2p^2q^2\overline{\varphi}_\mathbf{q}^2\right)^2+B\left(\lambda^t_{4M}\right)^3+(2B-4i\frac{\omega_*}{\tau}\eta q^2)p^2q^2\overline{\varphi}_\mathbf{q}^2{\lambda^t}_{4M}=0,
\end{equation}
where
\begin{equation}
    B=-(\lambda^{p+}_\mathbf{p}+\lambda^{p-}_\mathbf{p}+\lambda^{p+}_\mathbf{r}+\lambda^{p-}_\mathbf{r})=2(D_\mathbf{p}+D_\mathbf{r})+\frac{i\omega_*}{\tau}\left(\eta(q^2+2p^2)-2\right).
\end{equation}

At leading $\textit{O}(\overline{\varphi}_\mathbf{q}^2)$-order \eqref{RootPhiExpansion} yields the purely oscillating solutions
\begin{equation}
    \lambda_1^t = \pm \sqrt{2}ipq\overline{\varphi}_\mathbf{q},
\end{equation}
similar to the modes of \eqref{TemperatureKH}. In order to find the real part of these modes we then have to proceed to order $\textit{O}(\overline{\varphi}_\mathbf{q})$ since the $\textit{O}(\overline{\varphi}_\mathbf{q}^{3/2})$-part identically vanishes. At that order \eqref{RootPhiExpansion} yields
\begin{equation}
    \lambda_{1/2}^t=\pm\sqrt{\pm\frac{\eta\omega_* q^2}{\sqrt{2}\tau}}\sqrt{pq\overline{\varphi}_\mathbf{q}}.
\end{equation}
Combining these results, in the large $\overline{\varphi}_\mathbf{q}$-limit we therefore have the four solutions
\begin{equation}\label{UnstableTertiaryKHlike}
    \lambda_{4M}^{t\pm}=\sqrt{2}ipq\overline{\varphi}_\mathbf{q}\pm \sqrt{-\frac{\eta\omega_* q^2}{\sqrt{2}\tau}}\sqrt{pq\overline{\varphi}_\mathbf{q}},
\end{equation}
\begin{equation}\label{StableTertiaryKHlike}
    \lambda^{t}_{4M}=-\sqrt{2}ipq\overline{\varphi}_\mathbf{q}\pm \sqrt{\frac{\eta\omega_* q^2}{\sqrt{2}\tau}}\sqrt{pq\overline{\varphi}_\mathbf{q}},
\end{equation}
of which only the first is unstable since $\omega_*<0$. Do keep in mind that we will continue to use the \mbox{$+$-superscript} for the most unstable mode, regardless of whether $\overline{\varphi}_\mathbf{q}$ is large or not. 

Converting the solutions corresponding to \eqref{UnstableTertiaryKHlike} from $k$-space to real space using the zonal profile \eqref{4MZonalProfile} we find that they take the form
\begin{equation}
    \tilde{\varphi}=\left(1\pm\sqrt{2}\sin{qx}\right)\Re{\left(e^{ipy+\lambda^\pm_{4M}t}\right)}.
\end{equation}
It is apparent that the $x$-envelope, being given by the first factor, predominantly localises the unstable mode around minima of the zonal flow velocity and the stable mode around maxima, entirely in accordance with the picture that these points are tertiary (de-)stabilising as outlined in Section \ref{TertiaryInstabilityLocal}. Furthermore it is seen that, despite not being sufficiently localised for the treatment of Section \ref{TertiaryInstabilityLocal} to be justified, this result nevertheless agrees with the large $\overline{\varphi}$-limit of \eqref{TertiaryDispersionLocal} up to numerical constants.

Now turning to the opposite small $\overline{\varphi}_\mathbf{q}$-limit, we are interested in how the unstable primary mode is modified by the presence of a small zonal flow. Taylor expanding \eqref{TertiaryDispersion4M} around $\lambda^t_{4M}=\lambda^{p+}_\mathbf{p}$ one straightforwardly obtains the solution
\begin{equation}\label{SmallZonalTertiary}
    \lambda^{t+}_{4M}=\lambda^{p+}_\mathbf{p}-Cp^2q^2\overline{\varphi}_\mathbf{q}^2
\end{equation}
where
\begin{equation}\label{Cexpression}
    C=\frac{A(\lambda^{p+}_\mathbf{p})(\lambda^{p+}_\mathbf{p}-\lambda_\mathbf{r}^{p+})^\dag(\lambda^{p+}_\mathbf{p}-\lambda_\mathbf{r}^{p-})^\dag}{(\lambda^{p+}_\mathbf{p}-\lambda^{p-}_\mathbf{p})|(\lambda^{p+}_\mathbf{p}-\lambda^{p+}_\mathbf{r})(\lambda^{p+}_\mathbf{p}-\lambda^{p-}_\mathbf{r})|^2}.
\end{equation}

\begin{figure}
\centering
\includegraphics[width=0.75\linewidth]{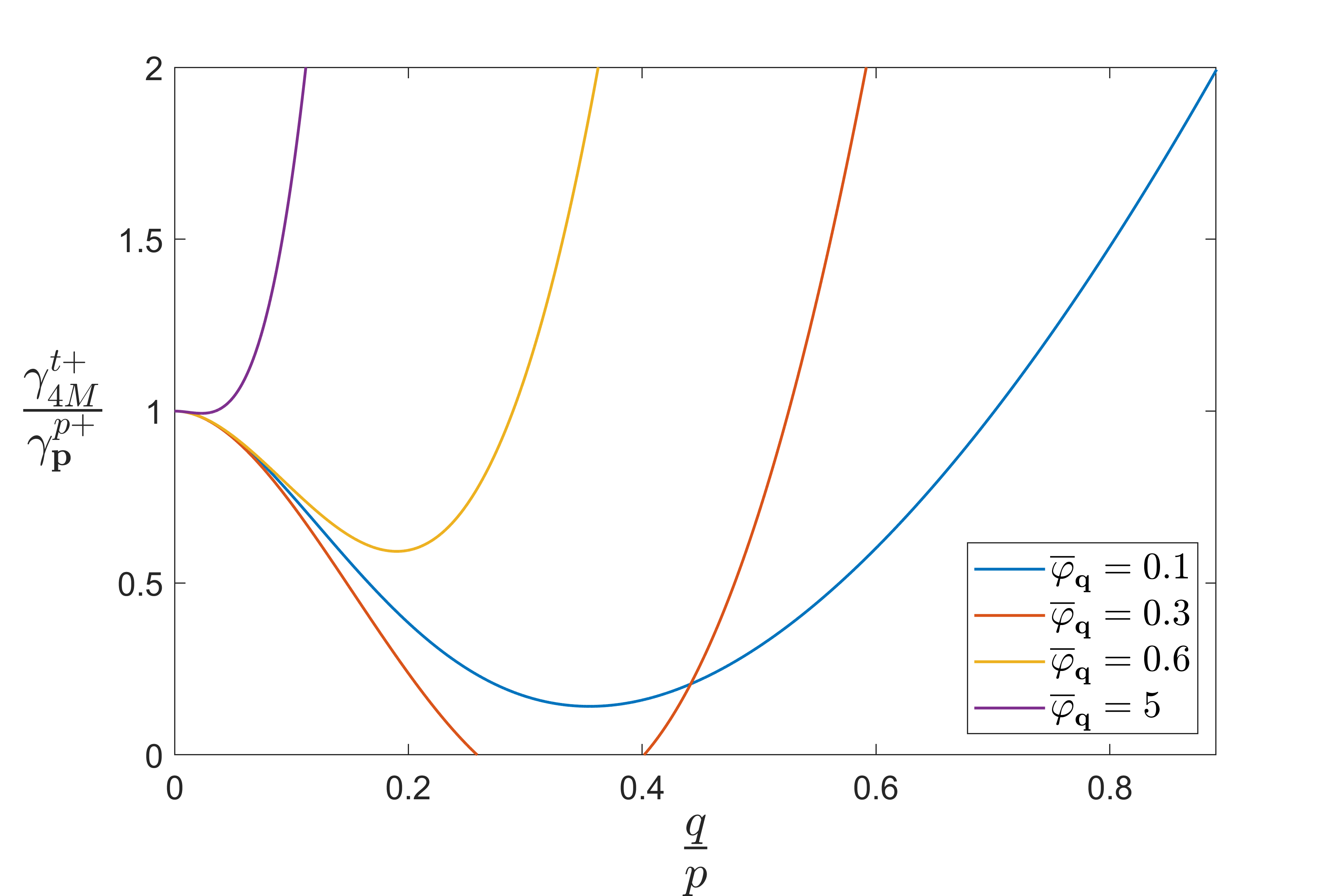}
\caption{4-mode tertiary growth rate $\gamma^{t+}_{4M}$ of the most unstable poloidal band satisfying \eqref{MaxUnstableky} for $(\omega_{d0},\omega_{*0},\tau,D,\eta)=(-0.8,-1.02,1,0.5,1.8)$ normalised by the unmodified primary growth rate $\gamma^{p+}_\mathbf{p}$, as a function of the zonal wavenumber $q$ for $\overline{\varphi}_\mathbf{q}=0.1$, 0.3, and 0.6 respectively}
\label{4MTertiaryFigure}
\end{figure}

Let us now employ a subsidiary ordering in $q$ to see how $\Re(C)$ behaves in the two limits $q^2\ll1$ (keeping $q^2\gg\overline{\varphi}_\mathbf{q}^2$) and $q^2\gg1$ (keeping $q^2\overline{\varphi}_\mathbf{q}^2\ll1$) to see whether the most unstable mode is initially stabilised or destabilised by the presence of zonal flows at these scales. Because it is apparent that the denominator of \eqref{Cexpression} is positive, and because the numerator of $C$ becomes
\begin{equation}\label{largeqC}
   -2\frac{\omega_*^2\eta^2q^4}{\tau^2}(|\lambda_\mathbf{p}^{p+}|^2+D_\mathbf{r}\gamma^{p+}_\mathbf{p}),
\end{equation}
for large $q$-values, it is apparent that $\Re({C})$ is negative for small scale zonal flows which therefore are destabilising already at small amplitude. As $\overline{\varphi}_\mathbf{q}$ is further increased we furthermore know that the mode under consideration transitions into \eqref{UnstableTertiaryKHlike}, and so we can conclude that small scale zonal flows are always destabilising. In fact it is numerically found that the transition to instability with increasing $q$ occurs much before this limit, already as $q\sim p$ as can be seen in Figure \ref{4MTertiaryFigure}. Note that this point is still much below that where the KH  -like $(qp\overline{\varphi}_\mathbf{q})^\alpha$-scaling of \eqref{UnstableTertiaryKHlike} develops, and thus the mode is still ostensibly ``more primary" in character.

If $q$ on the other hand is small, the numerator of $C$ becomes
\begin{equation}\label{SmallqC}
    16\left(\gamma_\mathbf{p}^{p+}\right)^2\left(i\frac{\omega_*}{2\tau}-\frac{\omega_*^2(1-\eta p^2)}{4\tau^2\gamma^{p+}_\mathbf{p}}\right)\eta q^2.
\end{equation}
This has a positive real part for those modes with $\eta p^2>1$, including the most unstable primary mode satisfying \eqref{MaxUnstableky}, and so large scales zonal flows are initially stabilising for these modes. It should be noted that, despite \eqref{SmallqC} reversing sign for modes of smaller $p$, the zonal flow does initially not destabilise large scale drift waves. As can be seen from the linear growth rate \eqref{LinearGrowth}, for these values of $p$ the small $q$ sidebands are in fact more unstable than the pure drift mode. Thus upon repeating the calculation above, but instead expanding around $\lambda^t_{4M}=\lambda_\mathbf{r}^{p+}$, one finds precisely the opposite stabilisation effect on the sidebands. These results can be summarised by the observation that, at small zonal amplitude, the tertiary mode constitutes a sort of weighted average of its constituent primary modes.

With the asymptotic behaviour of \eqref{UnstableTertiaryKHlike} and \eqref{SmallZonalTertiary} in hand, it is clear that the initial stabilisation \eqref{SmallZonalTertiary} of small amplitude zonal flows must reverse as the amplitude is increased, and there necessarily exists some zonal flow amplitude which is most stable. Precisely this can be seen in Figure \ref{4MTertiaryFigure}, where the most unstable 4M tertiary growth rate $\gamma_{4M}^{t+}$ for the example system $(\omega_{d0},\omega_{s0},\tau,D,\eta)=(-0.8,-1.02,1,0.5,1.8)$ with linear instability threshold $\eta^0=1.64$ (which will be the focus of the remainder of this paper), is plotted. In accordance with \eqref{largeqC} and \eqref{SmallqC}, as $\overline{\varphi}_\mathbf{q}$ begins to be increased the tertiary mode initially stabilises for $q\lesssim p$, and destabilises for $q$ of greater magnitude.

\subsection{Role of the Tertiary Instability for the Dimits transition}\label{RoleOfTheTertiaryInstability}

Extrapolating the consequences of the findings above to the dynamics of the zonally dominated states typical of the Dimits regime, some conclusions can be drawn. If zonal profile conditions are not ideal, said profile will fail to suppress the tertiary instability. Then drift waves will grow in amplitude to eventually affect the zonal profile. While such conditions prevail, the zonal profile will evolve through different configurations in a process we will refer to as \textit{zonal profile cycling}. In this process, energy will continue to be injected into the drift waves at a faster rate for more tertiary unstable zonal profiles. Thus the profile should be observed with higher probability in a state of low tertiary growth. Indeed, it is expected that a state of absolute tertiary stability could be sustained indefinitely. In conclusion we argue that the tertiary instability therefore preferentially selects a set of zonal profiles which will predominantly appear as the system evolves.

Because the zonal flows by construction are linearly undamped, a tertiary stable zonal profile can emerge that sustains the system in a state of suppressed turbulence, so long as the decaying residual drift wave activity, in turn, does not affect it too much. We will refer to such profiles as \textit{robustly stable}. Now, from the 4M-result above, we can extrapolate that the tertiary instability for our system exhibits only a finite ability to be stabilised. Naturally this means that the number of robustly stable zonal profiles should decrease as the driving gradient $\eta$ is increased. At some point, none remain, so turbulence and transport must arise. If this point indeed corresponds to the Dimits transition, then the only the relevant features needed to explain the Dimits shift should be the tertiary instability and the ability of the zonal flows to cycle through stabilising profiles. 

Of course it is possible that, even in the absence of collisional zonal damping, a stable zonal state cannot be attained. Another possibility is that some nonlinear mechanism continues to reduce transport above the tertiary instability threshold, making the Dimits threshold of appreciable transport not coincide with the tertiary threshold. An example of such a feature, already observed and explored in other systems, would be e.g. the ferdinons of \citet{Ivanov2020}. For our system however, this does not seem to be the case, and, as we will see, the tertiary instability alone seems sufficient to explain the Dimits shift. That is, below a certain point $\eta=\eta^{\mathrm{NL}}$, tertiary stable zonal profiles are always able to form and completely quench transport, while above it they cease to manifest and the time-averaged transport levels rapidly increase with $\eta$.

In conclusion, it should be noted that precisely what profiles are robust is a delicate and highly nontrivial question, which nevertheless ultimately decides when the Dimits transition occurs. Thus the tertiary instability should not enter Dimits shift picture via so simple a rule as ``the Dimits regime should end when the zonal amplitude becomes too large" as envisioned by \citet{Rogers2000b}, nor ``the Dimits regime should end when the zonal amplitude becomes too small" as \citet{Zhu2020} stated. Though the latter may hold when collisional damping limits the zonal amplitude, in general it is the much more nebulous question of ``can a robust zonal profile be reached and sustained during the subsequent transient period of decay" which must be answered.

\section{Nonlinear simulation results}\label{NonlinearSimulationsResults}

In order to thoroughly investigate the strongly driven system, it was first simulated pseudospectrally for several configurations on a square grid using a sixth order Runge-Kutta-Fehlberg method including 512x256 modes with $0\leq k_x,k_y\leq 5p_m$ and with dealiasing using the 3/2-rule. Sensitivity scans with regards to the number of modes and the minimum wavenumbers found this selection to be well beyond what was necessary for convergence within the Dimits regime, so long as the most unstable drift wave $p_m$ was included.

\begin{figure}
\centering
\includegraphics[width=0.9\linewidth]{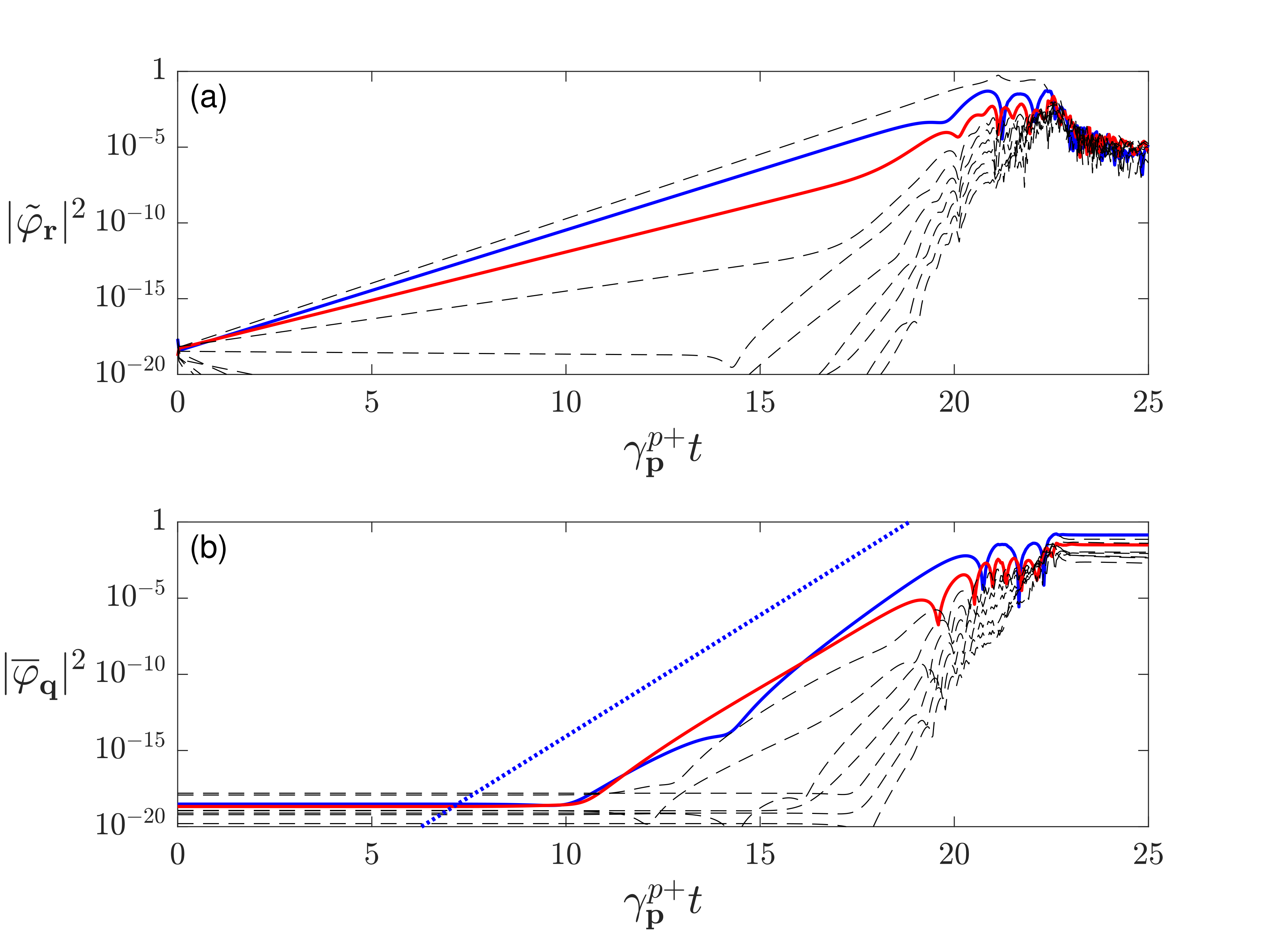}
\caption{(a) Drift wave and (b) zonal mode amplitudes during the initial growth phase for the system of Figure \ref{4MTertiaryFigure}. The blue line corresponds to the mode with smallest radial wavenumber $q_\mathrm{min}$, the red line to its second harmonic $2q_\mathrm{min}$, and other modes are denoted by black dashed lines. After an initial linear phase, the sideband-sideband-interaction of $\tilde{\varphi}_{(q_\mathrm{min},p)}$ excites $\overline{\varphi}_{(q_\mathrm{min},0)}$ and $\overline{\varphi}_{(2q_\mathrm{min},0)}$, which thus grow at a rate proportional to $\sim|\tilde{\varphi}_{(q_\mathrm{min},p)}|^2$, which is plotted with a dotted blue line for comparison. Modes of higher and higher $q$ are then excited one by one, until the zonal flows reach a magnitude comparable to the drift waves, which are then suppressed.}
\label{InitialGrowth}
\end{figure}

The nonlinear simulations usually employed a $\mathbf{k}$-independent $D_\mathbf{k}$, and were initiated with small Guassian noise of (normalised) energy density $10^{-8}$. As can be seen in Figure \ref{InitialGrowth}, the expected behaviour is then observed where primary modes emerge until they are strong enough to nonlinearly engage the zonal modes. In accordance with the secondary mode analysis of \ref{SecondaryInstability}, sideband-sideband interactions here play a vital role for initial zonal mode growth to occur, and so necessarily the second zonal harmonic, i.e. the mode with twice the smallest wavenumber $q_\mathrm{min}$, is primarily engaged. Following this, the largest scale zonal modes also begin to grow appreciably, until they in turn reach sufficient amplitude for nonlinear interactions to quickly shuffle energy from the unstable modes to higher and higher $q$-sidebands. These, in turn, engage higher $q$ zonal modes to affect the primary growth, and the growth phase ceases. This typically occurs when both the drift wave and zonal flow energy densities reach a comparable magnitude of around 1.

It is important to note that, as a consequence of there being no direct coupling between drift waves of differing poloidal wavenumber, the system typically stratifies into separate $p$-layers only interacting with each other via their influence on the zonal flow. In the Dimits regime, where necessarily $D_\mathbf{p}\sim\gamma_\mathbf{p}^{p+}$, only those few modes with $p$ around $p_m$, satisfying \eqref{MaxUnstableky}, are linearly unstable. Consequently, the layer corresponding to the dominant primary mode becomes solely dynamically important, as is borne out in simulations. Though the zonal profile could excite other bands through the tertiary instability, since the primary band is the most tertiary unstable this is not borne out in practice. More layers become important only once they become primary unstable at larger $\eta$-values, occurring above the point at which the transition to continuous transport occurs.

\begin{figure}
\centering
\includegraphics[width=0.7\linewidth]{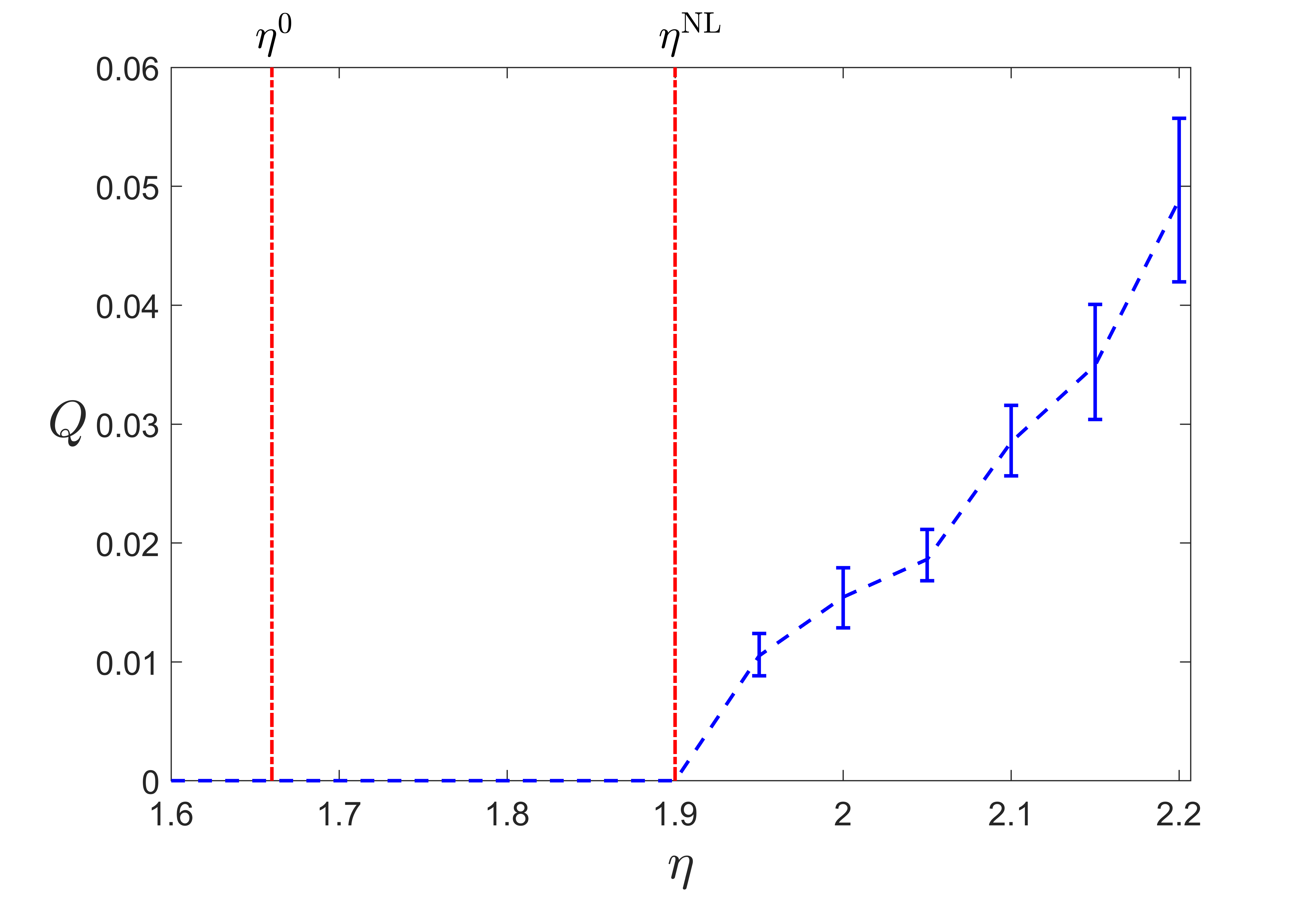}
\caption{Long time-averaged heat flux, given by \eqref{HeatFlux}, as a function of $\eta$ for $(\omega_{d0},\omega_{s0},\tau,D)=(-0.8,-1.02,1,0.5)$. The linear instability threshold occurs at $\eta^0\approx1.66$, yet finite heat flux only commences beyond $\eta^{\mathrm{NL}}\approx1.9$, constituting a clear Dimits shift. Between these points, the system relaxes to completely stable purely zonal states.}
\label{HeatTransport}
\end{figure}

Now, initial saturation amplitudes of both zonal and drift waves exhibit a very slight dependence on the initialisation amplitude. This is because differing primary/sideband growth rates causes there to be more or less initial energy to distribute, depending on the amount of time the primary mode has had to grow before the sidebands trigger zonal growth. Nevertheless, in the Dimits regime the zonal amplitude usually quickly returns to a system-configuration-dependant typical amplitude, like the tertiary analysis of \ref{TertiaryInstability4M} suggests. Once there, a rectangular zonal shear profile resembling a less developed version of the staircase states observed by e.g. \citet{Dif-Pradalier2010}, \citet{Kobayashi2012} or \cite{Peeters2016}, quickly develops, which can then efficiently suppress the amplitude of drift waves, typically on the order of 2 magnitudes (but occasionally much more). All that then remains are the localised tertiary modes which will eventually die if stable or grow back if unstable. Observing the heat flux after a long time has passed, like in Figure \ref{HeatTransport}, the presence of a Dimits shift is thus revealed, since stable states only exist close to the linear threshold $\eta^0$.

\subsection{Drift Wave Bursts}\label{DriftWaveBursts}

\begin{figure}
\centering
\includegraphics[width=0.83\linewidth]{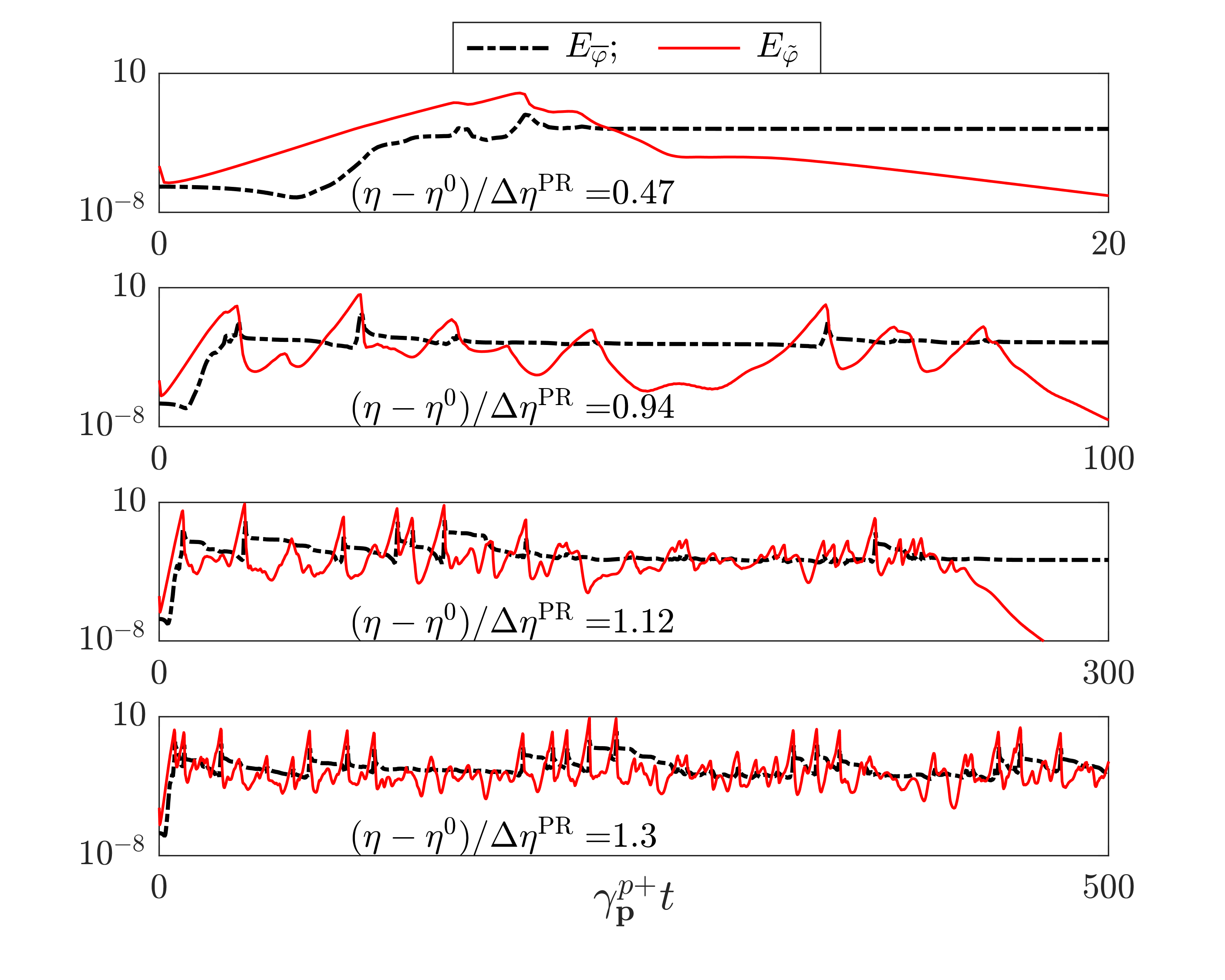}
\caption{Time evolution for the configuration of Figure \ref{HeatTransport} of the zonal (black dashed) and nonzonal (red) energy densities $E_{\overline{\varphi}}$ and $E_{\tilde{\varphi}}$, given by \eqref{EnergyDefinition}, with increasing instability $\eta$ above the linear threshold $\eta^0$, where $\eta^0$ is the linear instability threshold and $\Delta\eta^{\mathrm{PR}}$ is the predicted Dimits transition as introduced in Section \ref{ReducedModeDimitsShiftEstimate}. As the system becomes more unstable it is observed to take longer to arrive at a completely stable zonal state, while simultaneously exhibiting more rapid bursty behaviour.}
\label{FreeEnergyEvolution}
\end{figure}

As can be seen Figure \ref{FreeEnergyEvolution}, when the instability parameter $\eta$ is increased away from the linear threshold $\eta^0$ while other parameters are kept the same, the initial zonal profiles attained are commonly completely tertiary stable. However as $\eta$ is further increased this usually ceases to be the case, since the primary/sideband coupling then fails to remain strong enough for the primary mode to decay together with the damped sidebands. Consequently, a spreading turbulence burst destroys the initial zonal state, cycling through zonal profiles until another stable state is reached. The rapidity by which such bursts occur, and the time until a stable zonal profile is attained, both rapidly increase with $\eta$, unless the cycling by happenstance quickly produces a stable state. At even larger $\eta$-values, no stable state is ever attained. Furthermore, the typical burst amplitude is also reduced as a result of less efficient drift wave quenching. 

Note that this entire burst pattern is, on the surface, similar to the zonal/drift predator-prey interactions commonly observed in many systems as a result of zonal damping \citep[see e.g.][]{Malkov2001,Kobayashi2012}. However, it differs fundamentally in that the turbulent bursts are not typically accompanied by large zonal amplitude swings. Instead, it traces its origin to tertiary mode localisation.

\begin{figure}
\centering
\includegraphics[width=\linewidth]{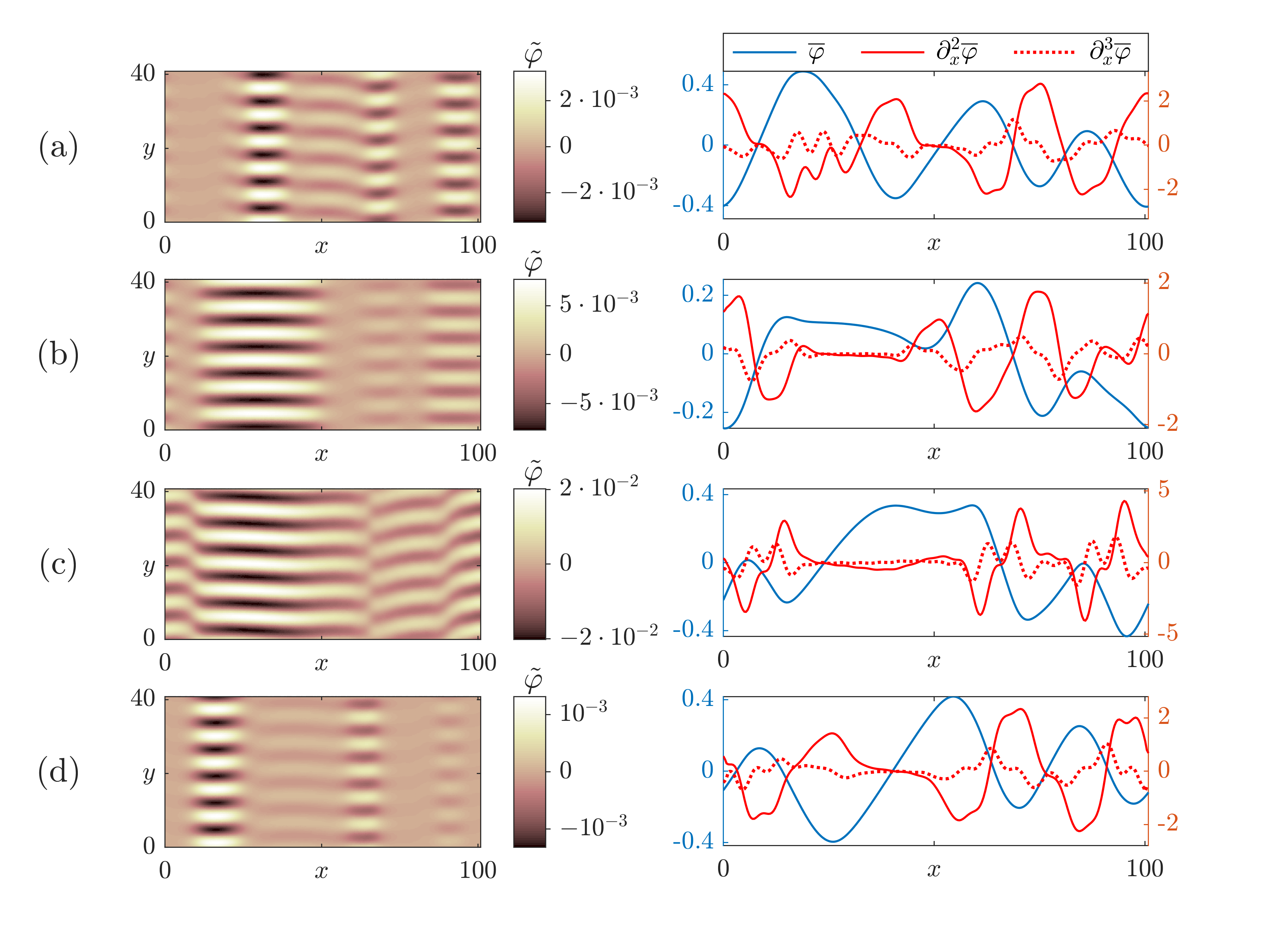}
\caption{Snapshots for the system of Figure \ref{4MTertiaryFigure} at (a) $\gamma^{p+}_\mathbf{p}t=33$, (b) $\gamma^{p+}_\mathbf{p}t=35.8$, (c) $\gamma^{p+}_\mathbf{p}t=37.5$, and (d) $\gamma^{p+}_\mathbf{p}t=47$ depicting drift waves on the left and the zonal potential (blue), the zonal flow shear $\partial_x^2\overline{\varphi}$ (red), and its derivative $\partial_x^3\overline{\varphi}$ (dotted red) on the right. These depict a turbulent burst originating as an unstable tertiary mode at $x\approx 32$ where $\partial_x^2\overline{\varphi}=0$ and $\partial_x^3\overline{\varphi}>0$ at $\gamma_\mathbf{p}^{p+} t=33$, broadening and growing in amplitude between two tertiary unstable propagating zonal fronts at $\gamma_\mathbf{p}^{p+} t=35.8$ until the drift waves encompass the whole volume at $\gamma_\mathbf{p}^{p+} t=37.5$, rapidly modifying the zonal profile until a new zonally dominated state can be reinstated at $\gamma_\mathbf{p}^{p+}t=47$, but which exhibits seeded tertiary modes at $x\approx17$ and $x\approx64$ which will eventually repeat this process.}
\label{StaircaseSnapshots}
\end{figure}

To see how this is the case, some snapshots of a typical burst are displayed in Figure \ref{StaircaseSnapshots}. A zonal profile exhibits tertiary modes predominantly localised at the points where the conditions $\partial_x^2\overline{\varphi}=0$, and $\partial_x^3\overline{\varphi}>0$ are satisfied, in accordance with \eqref{TertiaryDispersionLocal}. Eventually the one at $x\approx32$ grows enough to affect the zonal profile at this point, resulting in a central flattening of the zonal amplitude. While the zonal shearing rate $\partial_x^2\overline{\varphi}$ remains 0 in the process, $\partial_x^3\overline{\varphi}$ is reduced, except for at the boundary of the full mode, causing a central tertiary instability reduction. The tertiary mode now becomes more unstable at its boundary, where the condition $\partial_x^3\overline{\varphi}>0$ is still maintained. As a result, the zonal flattening continues and the tertiary mode broadens behind a propagating zonal front, eventually encompassing much of the domain and destroying the zonal profile. 

After a period of zonal profile cycling, a stable zonal profile is eventually reestablished, which again quenches the drift waves down to the original amplitude. Typically the tertiary instability is now localised to different points, where seeded drift waves can eventually repeat the process. However, since these points were initially tertiary stable it takes a long time for a localised mode to fully develop. This explains the large swings in transport levels at marginally unstable $\eta$-values observed in Figure \ref{FreeEnergyEvolution}.

As a final remark it is worth noting that during an entire typical burst process, the box averaged zonal shear magnitude $|\partial_x^2\overline{\varphi}|$ remains comparable to the primary growth rate $\gamma^{p+}_\mathbf{p}$. This is a typical result also obtained in previous investigations of zonally dominated states \citep{Waltz1998,Kinsey2005,Kobayashi2012} known as the quench rule.

\section{Reduced mode Dimits shift estimate}\label{ReducedModeDimitsShiftEstimate}

Having identified the importance of the tertiary instability for the Dimits transition in Sections \ref{TertiaryInstabilityLocal}, \ref{TertiaryInstability4M}, and \ref{NonlinearSimulationsResults}, we now turn our attention to the problem of attempting to predict the size of the Dimits shift using tertiary instability analysis. For the system under consideration it is clearly necessary that there exists tertiary stable zonal profiles for the system to be located within the Dimits regime. This has been exploited before by \citet{Zhu2020,Zhu2020a} in other systems where full stability characterises the Dimits regime to match the Dimits shift threshold to a tertiary transition. However this matching could only be done in hindsight by tuning a representative zonal curvature value by hand, and did thus not constitute a prediction.

To predict the Dimits shift the major problem one encounters is, as we outlined in Section \ref{RoleOfTheTertiaryInstability}, the full multitude of possible zonal profiles, and accounting for how these can be generated through nonlinear interactions with the drift waves. Some profiles may fail to form nonlinearly, while others fail to be robustly stable enough to persist while the residual drift waves decay. Thus it may not be sufficient that a tertiary stable zonal profile exists for the system to be in the Dimits regime. Thoroughly accounting for this seems a herculean task and instead some major simplifying assumptions have to be employed.

An example of a simple method doing just that is the heuristic prediction of \citet{St-Onge2018}, which relies on the same tertiary 4-mode Galerkin truncation as \eqref{4MGalerkin} in lieu of accounting for the complex interplay of the many modes constituting the full zonal flow profile. In a modified Terry-Horton system it was postulated that a typical mode could be approximated by that maximally coupled tertiary 4M satisfying the condition $\lambda^{t+}_{4M}=\lambda^{t-}_{4M}$ (in our notation). Approximating the coupling of the primary mode to its sidebands through the 4M-interaction alone, and assuming that it is the most unstable primary mode that determines when the Dimits regime ends, the Dimits transition was then taken to occur when this cluster of maximally coupled modes became unstable. 

\citet{St-Onge2018} found this prediction to be in excellent agreement with the observed Dimits transition. However the sensitivity of this transition to numerical dissipation, the small transport levels immediately beyond this point, and the slow evolution made it somewhat difficult to definitively classify a state as stable or turbulent close to this threshold in subsequent reproductions of this system by \citet{Zhu2020a}. On the other hand, for our purposes the major flaw of this prediction is the fact that, with the inclusion of 3 modes for each $\mathbf{k}$, there ceases to form anything resembling maximally coupled modes. Nevertheless, because of the simplicity of such a scheme we now look for a similar zonal profile reduction to arrive at a reduced mode prediction.

The first key feature of the present system when attempting to arrive at a simplified prediction is the aforementioned stratification observed in the nonlinear simulations. As mentioned, it is the poloidal band corresponding to the most unstable primary mode $\mathbf{p}$ which goes tertiary unstable first, and thus solely determines when the Dimits regime ends. Secondly we recall the result of Sections \ref{TertiaryInstability4M} and \ref{RoleOfTheTertiaryInstability} that the tertiary instability frequently acts in such a way as to push the zonal amplitude $\overline{\varphi}_\mathbf{q}$ towards its most stabilising value. At least as long as the 4M-interaction is dominant, the zonal profile should therefore repeatedly evolve into a state similar to this one.

For the final piece of the puzzle a much more reductive simplification is employed which relegates this method firmly into being an estimate. We choose to approximate a typical full zonal profile with a single mode, with wavelength $q$, which is the most 4M tertiary stabilising. Of course any real profile will include many more modes of varying amplitude. Nevertheless, some of these will act to destabilise and some to stabilise the tertiary modes, and so we assume that their cumulative effect can be approximated by a representative mode. Indeed the surprising qualitative similarity between the 4M tertiary growth rate in the high $\overline{\varphi}$-limit \eqref{UnstableTertiaryKHlike} and the local tertiary \eqref{TertiaryDispersionLocal} hints that this approximation may be less far-fetched than it seems.

\begin{figure}
\centering
\includegraphics[width=0.85\linewidth]{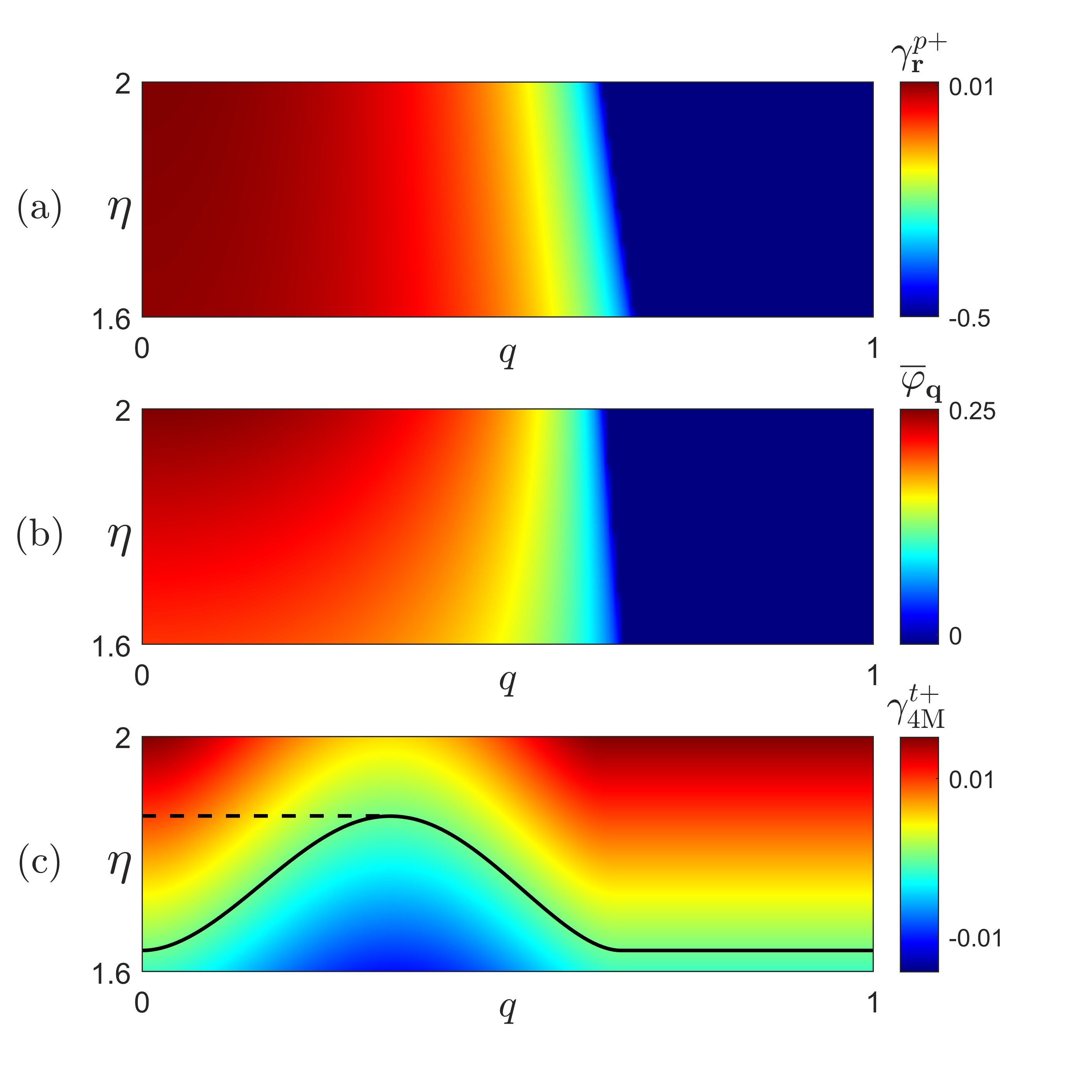}
\caption{(a) The primary instability growth rate $\gamma^{p+}_\mathbf{r}$, as given by \eqref{LinearGrowth}, for the sideband modes $\mathbf{r}^{\pm}$ of the 4M system \eqref{4MGalerkin} with poloidal wavenumber \eqref{MaxUnstableky}, (b) the most stabilising zonal amplitude $\overline{\varphi}_\mathbf{q}$ as given by \eqref{cond2}, (c) the corresponding 4M-tertiary growth rate $\gamma^{t+}_{4M}$ as a function of $\eta$ and $q$, all for the configuration of Figure \ref{HeatTransport}. The resulting 4M-tertiary Dimits prediction $\eta^{\mathrm{PR}}$ is indicated by a dashed line, constituting a predicted Dimits shift $\Delta\eta^\mathrm{PR}=\eta^{\mathrm{PR}}-\eta^0$ of $\sim0.23$.}
\label{DimitsEstimate}
\end{figure}

Combining these pieces, we conclude that the Dimits transition should occur at around that point $\eta^{\mathrm{PR}}$ where our single zonal mode ceases to be able to stabilise the most unstable primary mode, leaving no robustly stable zonal state attainable. Thus we can express our Dimits shift prediction $\Delta\eta^{\mathrm{PR}}$ as
\begin{equation}\label{PredictionEquation}
    \Delta\eta^{\mathrm{PR}}=\eta^{\mathrm{PR}}-\eta^0,
\end{equation}
where $\eta^{\mathrm{PR}}$ is the solution to the constrained optimisation problem
\begin{equation}\label{cond1}
    \frac{\partial \gamma^{p+}_\mathbf{p}}{\partial p}=0,
\end{equation}
\begin{equation}\label{cond2}
    \frac{\partial\gamma^{t+}_{4M}}{\partial |\overline{\varphi}_\mathbf{q}|}=0,
\end{equation}
\begin{equation}\label{cond3}
    \frac{\partial \gamma^{t+}_{4M}}{\partial q}=0,
\end{equation}
\begin{equation}\label{cond4}
    \gamma^{t+}_{4M}(\eta^{\mathrm{PR}}) = 0.
\end{equation}
In Figure \ref{DimitsEstimate} this method can be seen in action for the configuration of Section \ref{NonlinearSimulationsResults}. Though we again stress that \eqref{PredictionEquation} is clearly a nonrigorous estimate of the Dimits transition, the broader accuracy of which has to be confirmed by comparison with nonlinear simulations we will perform in Section \ref{PredictionAndNonlinearComparison}, this predicted Dimits transition at $\eta^{\mathrm{PR}}=1.86$, corresponding to a Dimits shift of $\Delta\eta^\mathrm{PR}\approx0.23$, is indeed close to the point $\eta^{\mathrm{NL}}\approx1.9$ of Figure \ref{HeatTransport} where the drift waves observed in nonlinear simulations fail to vanish.

Now, in principle it should be possible to apply the estimation method as just outlined to other systems, up to and including gyrokinetics, so long as one is mindful of what zonal profiles are typically observed and whether some slight modification is necessary to account for these. However, it can only be expected to be useful so long as collisional damping is low enough for the Dimits regime to be characterised by sufficiently stable zonal states, so that the Dimits transition coincides with the point of tertiary destabilisation. Should this not be the case, some other method would have to be employed such as e.g. that of \citet{Ivanov2020}, asymptotically accurate in the highly collisional limit, which investigates whether the effect of the turbulence upon the zonal flow either counteracts or reinforces its collisional dissipation.

\subsection{Comparison of Prediction and Nonlinear Results}\label{PredictionAndNonlinearComparison}

The question now is to what extent the prediction as just outlined in Section \ref{ReducedModeDimitsShiftEstimate} is generally accurate. To investigate this question we are greatly aided by the observed poloidal stratification and how the dominant primary band is the most tertiary unstable. This means that we can restrict ourselves to only include zonal modes and the most unstable poloidal band in nonlinear simulations, which constitutes a tremendous speedup enabling us to investigate a very wide range of different configurations. This reduction was found to have no appreciable effect on the observed Dimits shift for any of the many disparate test cases, and seems to be uniformly valid for this investigation.

The specific way in which the Dimits shift for a configuration with a given $(\omega_{d0}, \omega_{*0}, \tau,\eta^0)$ was determined can be described as follows. A set of simulations with increasing $\eta$ were performed and were allowed to run continuously until a fully stabilising profile was obtained and all drift-wave amplitudes died down below their original values. If this had not occurred within the time $t_\mathrm{end}=3000\gamma^{-1}$, the simulation was stopped, and the Dimits transition taken to be the final $\eta$ where a stable state arose. 

Across thousands of simulations and nearly a hundred configurations, only a handful of times did a simulation with a higher $\eta$ reach stability while one below it failed to do so, and in the few cases for which this occurred all were located right at the Dimits threshold. As outlined in Section \ref{RoleOfTheTertiaryInstability}, this is because the space of robustly stable zonal configurations rapidly shrinks to 0 at the true Dimits threshold $\eta^{\mathrm{Di}}$, beyond which no stable states exist. Thus, although it occasionally happens that a stable profile is quickly obtained, the average time $t_\mathrm{avg}$ to attain one of these stable states in nonlinear simulations diverges at $\eta^{\mathrm{Di}}$,
\begin{equation}
    \lim_{\eta \to \eta^{\mathrm{Di}}}t_{\mathrm{avg}}=\infty.
\end{equation}
Therefore the observed Dimits transition $\eta^{\mathrm{NL}}$, obtained in nonlinear simulations, typically varies very little with respect to $t_\mathrm{end}$ once it is sufficiently large,
\begin{equation}
    \lim_{t_{\mathrm{\mathrm{end}}}\to \infty}\frac{\partial \eta^{\mathrm{NL}}}{\partial t_\mathrm{end}}=0.
\end{equation}
With these observations we feel justified in claiming that this method of determining the Dimits shift is robust for our system.

\begin{figure}
\centering
\includegraphics[width=\linewidth]{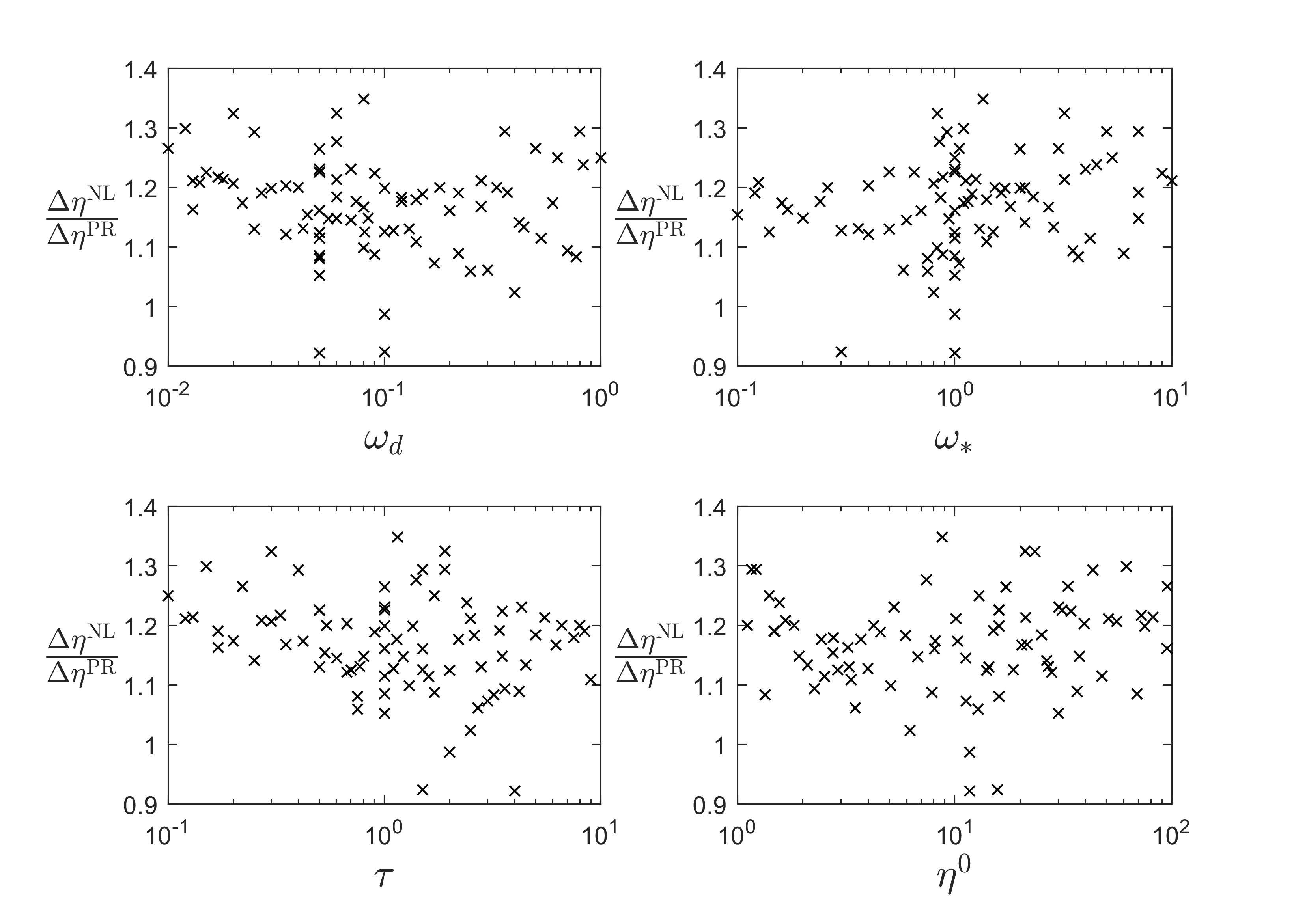}
\caption{Dimits shift mismatch between the reduced mode prediction $\Delta\eta^{\mathrm{PR}}$ of Section \ref{ReducedModeDimitsShiftEstimate} and what is observed in nonlinear simulations $\Delta\eta^{\mathrm{NL}}$ as a function of the configuration parameters $\omega_d$, $\omega_*$, $\tau$, and $\eta^0$ for a set of configurations where all parameters were simultaneously randomly chosen. The prediction is generally seen to underpredict the actual shift by some 5-30\% but otherwise remain consistent across configurations.}
\label{CombinedDimitsResults}
\end{figure}
A comparison of the predicted Dimits shift $\Delta\eta^{\mathrm{PR}}$ and the nonlinearly obtained Dimits shift $\Delta\eta^{\mathrm{NL}}$ for configurations within the wide range given by $\omega_{d0}\in[-10,-10^{-5}]$, $(\omega_{*0},\tau)\in[-10,-10^{-2}]$, $\eta^0\in[10^{-0.5},10^{2.5}]$, where all these parameters have been simultaneously randomly chosen, can be seen in Figure \ref{CombinedDimitsResults}. The prediction is seen to generally underpredict the Dimits shift by about 15\% barring a few outliers varying by some 10\%. Nevertheless this is a favourable result because of how very consistent it remains across such a wide range of different configurations, while ultimately being so simple. Indeed no trend for the deviation with respect to any parameter can be discerned.

Finally, as mentioned in Section \ref{TheDampingOperator}, the nonlinear findings presented in Section \ref{NonlinearSimulationsResults}, and which were used to check the accuracy of prediction \eqref{PredictionEquation}, used a constant (with respect to $k_y$) dissipation $D_\mathbf{k}$. In principle the validity of these findings could cease to hold for other types of damping, since its inclusion would modify both the prediction and nonlinear Dimits transition slightly. Nevertheless this seems to be an unimportant effect in so far as the matching between the two, just like for \citet{St-Onge2018}, appears to remain mostly intact for different reasonable hyperviscosities, as observed across multiple exploratory simulations. Noting this, we feel content that using a constant $D_\mathbf{k}$ is general enough for our purposes and choose not to investigate this in any further detail.

\section{Discussion}\label{Discussion}

What has been demonstrated in this paper is that, through an appropriate asymptotic expansion, the moment hierarchy of gyrokinetics can self-consistently be closed at second order, resulting in a strongly driven gyrofluid system with both linear drive and nonlinear drift/zonal interactions. With the \textit{ad hoc} introduction of additional damping, meant to encapsulate the effect of Landau damping, this system can be made to exhibit a Dimits shift. 

Within the system studied in this paper, the Dimits transition was found to correspond to that point at which the zonal profile can no longer form a robustly tertiary stable profile to eventually kill all drift waves. This is an entirely different mechanism to the one found by \citet{Ivanov2020} in a very similar system. The difference arises because in that system, being fundamentally collisional, zonal flows decay and no self-stable zonal flows exist. Instead the Dimits regime is characterised by zonal staircases continually rebuilt by momentum influx from small drift wave turbulence, yielding low mean transport rates. At high collisionality \citet{Ivanov2020} could accurately predict the Dimits transition as that point where momentum flux turned destructive, but this scheme failed at lower collisionality. Likewise, were we to include zonal dissipation in our model our scheme would increasingly fail with increasing collisionality, since it relies on stable zonal states. Thus collisionality is clearly both a qualitatively and quantitatively important parameter to consider when one attempts to study the Dimits regime. 

Though similar, the transport bursts observed around the Dimits transition in this system are not strictly speaking the well-known zonal-drift wave predator-prey oscillations observed by e.g. \citet{Malkov2001} or \citet{Kobayashi2015a}, since the zonal energy typically only varies slightly during a drift wave burst. Furthermore, due to the lack of drift wave self-interactions it is clear that these bursts do not arise as a result of ferdinons \citep{VanWyk2016}. Instead it seems that these bursts trace their origin to the movement of tertiary unstable points as the zonal profile is modified, combined with the fact that new localised tertiary modes take so long to emerge, being close to marginal stability within the Dimits regime.

Similarly to the prediction of \citet{St-Onge2018}, it is possible to employ a consistently and comparatively accurate estimate of when the Dimits transition should occur by approximating the full zonal profile with a single mode, if it is taken to be that mode which is most stabilising. Naturally the validity of such a simplification may not in general be taken for granted; one has to be reasonably sure that the system under consideration has small enough collisional damping and that other kinds of nonlinear behaviour do not dominate during the Dimits transition so that the tertiary instability is still of prime importance. Since $\mathbf{E}\times \mathbf{B}$-shearing by strong zonal flows should remain the dominant nonlinear interaction, however, it seems likely that the Dimits transition continues to coincide with a tertiary stability threshold. Then one might, in more general collisionless systems, be able to approximate the typical zonal profiles with some reduced mode scheme, adapting equations \eqref{cond1}-\eqref{cond4} in some simple way, to maintain that computational simplicity that would make our theoretical method of Dimits shift estimation a practically useful and predictive tool. Because of this, future work aims to investigate whether this state of affairs holds for fully gyrokinetic systems. 

\section{Acknowledgements}

This work has been carried out within the framework of the EUROfusion consortium and has received funding from the Euratom research and training programme 2014-2018 and 2019-2020 under grant agreement No 633053. The views and opinions expressed herein do not necessarily reflect those of the European Commission.

\appendix

\section{Ordering Scheme and derivation} \label{Derivation}

We will here present the derivation giving rise to equations \eqref{phi_drift}-\eqref{t_par} which are the focus of this paper. Our starting point, repeated here for convenience, is the Fourier space dimensionless gyrokinetic equation
\begin{equation}\label{appendixgyrokineticequation}
    \left[ \frac{\partial}{\partial t} + i \omega_d \left( w_\parallel^2 + \frac{ w_\perp^2 }{ 2} \right) \right] h_{\mathbf{k}} + \PB{\Phi}{h}_\mathbf{k} = \left[ \frac{\partial}{\partial t} + i \omega_* \left( 1 + \eta \left( w^2-\frac{3}{2} \right) \right) \right] \Phi_\mathbf{k} f_0
\end{equation}
and the quasineutrality condition
\begin{equation}\label{appendixquasineutrality}
    \int d^3w J_0 h_\mathbf{k} = n (1 + \tau \hat{\alpha}) \varphi_\mathbf{k},
\end{equation}
with the conventions outlined in chapter \ref{BasicModel}. To these equations we we will now apply a subsidiary multiscale expansion on top of the fundamental gyrokinetic ordering via the introduction of a small parameter $\delta$. We start by specifying that
\begin{equation}\label{appendixOrdering}
    \frac{\omega_d}{\omega_*} \sim k_\bot^2 \sim \frac{1}{\eta} \sim \delta \ll 1
\end{equation}
be satisfied, while fixing the linear and nonlinear time scales $\omega_{L}$ and $\omega_{NL}$ according to
\begin{equation}\label{appendixomegas}
    \omega_* \sim \omega_{L} \sim \omega_{NL},
\end{equation}
to be able to describe both dynamics. 

In order to arrive at a closed fluid system from the fluid hierarchy we then expand the nonzonal components $\tilde{\varphi}_\mathbf{k}$ and $\tilde{h}_\mathbf{k}$ in orders of $\delta$ like
\begin{equation}
    \tilde{\varphi}_\mathbf{k} = \tilde{\varphi}_{\mathbf{k}0} + \tilde{\varphi}_{\mathbf{k}1} + \textit{O}(\delta^2\tilde{\varphi}_{\mathbf{k}0}), \;\;\; \tilde{h}_\mathbf{k} = \tilde{h}_{\mathbf{k}-1} + \tilde{h}_{\mathbf{k}0} + \textit{O}(\delta^2\tilde{h}_{\mathbf{k}-1}),
\end{equation}
with corresponding real space quantities $\varphi_0$, $h_0$, and so on, after which we likewise expand the zonal components $\overline{\varphi}_\mathbf{k}$ and $\overline{h}_\mathbf{k}$ like
\begin{equation}
    \overline{\varphi}_\mathbf{k} = \overline{\varphi}_{\mathbf{k}0} + \overline{\varphi}_{\mathbf{k}1} + \textit{O}(\delta^2\overline{\varphi}_{\mathbf{k}0}), \;\;\; \overline{h}_\mathbf{k} = \overline{h}_{\mathbf{k}0} + \overline{h}_{\mathbf{k}1} + \textit{O}(\delta^2\overline{h}_{\mathbf{k}}).
\end{equation}
Here, the notation is chosen such that the ordering
\begin{equation}\label{appendixhphi}
    \int d^3wv^i\tilde{h}_{\mathbf{k}j}\sim \tilde{\varphi}_{\mathbf{k}j}\;\;\;\mathrm{and}\;\;\;\int d^3wv^i\overline{h}_{\mathbf{k}j}\sim \overline{\varphi}_{\mathbf{k}j}
\end{equation}
is assumed to hold for arbitrary $i$ and $j$. However, perhaps counter-intuitively, we additionally order $\tilde{\varphi}_i \nsim \overline{\varphi}_i$ and $\tilde{h}_i \nsim \overline{h}_i$. This is permissible because the linear terms in \eqref{appendixgyrokineticequation} only affect the nonzonal components and because the nonlinear terms affecting zonal/nonzonal components are different, as we soon will see. Explicitly we will thus employ the ordering
\begin{equation}\label{appendixzonal/nonzonal}
    \frac{\tilde{\varphi}^2_i}{\overline{\varphi}^2_i} \sim \delta\;\;\;\mathrm{and}\;\;\;\frac{\tilde{h}^2_i}{\overline{h}^2_i} \sim \delta,
\end{equation}
which we eventually will see allows the exact closure we are aiming for, without giving rise to any contradiction. 

Now, the reason why the $\tilde{h}_{\mathbf{k}-1}$-term is present in the absence of a corresponding $\tilde{\varphi}_{\mathbf{k}-1}$-term merits some discussion, but essentially its inclusion is necessary because the dominant $\eta$-term in \eqref{appendixgyrokineticequation}, though possessing a vanishing density moment, has a nonvanishing temperature moment. To see this in action, we note that, using \eqref{appendixomegas}, the lowest order nonzonal perpendicular temperature moment of the gyrokinetic equation \eqref{appendixgyrokineticequation} is given by
\begin{equation}\label{ModifiedMomentEquation}
    \int d^3w w_\bot^2 \frac{\partial \tilde{h}_{\mathbf{k}-1}}{\partial t} + \int d^3w w_\bot^2 \PB{\overline{\varphi}_0}{\tilde{h}_{-1}}_\mathbf{k} = i \omega_* \eta \varphi_{\mathbf{k}0} \int d^3w w_\bot^2 \left( w^2 - \frac{3}{2} \right)f_0,
\end{equation}
and likewise for the parallel temperature moment, since $\Phi_\mathbf{k}=J_0\varphi_\mathbf{k}$ and
\begin{equation}\label{appendixJ0}
    J_0 = 1 - \frac{k_\bot^2 w_\bot^2}{2} + \textit{O}(\delta^2).
\end{equation}
Plainly the inclusion of the first $\tilde{h}_{\mathbf{k}-1}$-term is necessary to allow the last $\eta\tilde{\varphi}_{\mathbf{k}0}$-term in \eqref{ModifiedMomentEquation} to be linearly balanced. 

Since the linear terms, being proportional to $k_y$, vanish for zonal fields, these are in contrast to their equivalent nonzonal fields only driven by the nonlinear term
\begin{equation}\label{appendixTseparation}
    \int d^3w w_{\bot}^2 \PB{\tilde{\varphi}_0}{\tilde{h}_{-1}}_\mathbf{k}
\end{equation}
due to the fact that $\overline{\PB{\tilde{a}}{\overline{b}}}_\mathbf{k}=0$. Clearly, \eqref{appendixTseparation} is an order $\delta^{1/2}$ smaller than the terms of \eqref{ModifiedMomentEquation}, consistent with the explicit ordering separation of \eqref{appendixzonal/nonzonal}. This will mean that the zonal temperature moments fail to be dynamically relevant, and so we can restrict our attention to the former fields. Introducing the notation
\begin{equation}\label{Tnotation}
    \tilde{T}_{\bot\mathbf{k}}=\frac{1}{2n}\int d^3ww_\bot^2\tilde{h}_{\mathbf{k}-1},\;\;\; \tilde{T}_{\parallel\mathbf{k}}=\frac{1}{2n}\int d^3ww_\parallel^2\tilde{h}_{\mathbf{k}-1}
\end{equation}
we thus arrive at the relevant temperature equations
\begin{equation}\label{appendixTperp}
    \frac{\partial \tilde{T}_{\perp\mathbf{k}}}{\partial t} + \PB{\overline{\varphi}_0}{\tilde{T}_\perp}_\mathbf{k} = \frac{i \omega_* \eta \tilde{\varphi}_{\mathbf{k}0}}{2}
\end{equation}
and
\begin{equation}\label{appendixTpar}
    \frac{\partial \tilde{T}_{\parallel\mathbf{k}}}{\partial t} + \PB{\overline{\varphi}_0}{\tilde{T}_\parallel}_\mathbf{k} = \frac{i \omega_* \eta \tilde{\varphi}_{\mathbf{k}0}}{4},
\end{equation}
where the Maxwellian nature of $f_0$ has been used.

To determine the evolution of $\tilde{\varphi}_{\mathbf{k}0}$ and $\overline{\varphi}_{\mathbf{k}0}$ we now want to return to the gyrokinetic equation \eqref{appendixgyrokineticequation}. First, however, we note that the lowest order quasineutrality condition \eqref{appendixquasineutrality} for nonzonal components, using \eqref{appendixhphi} and \eqref{appendixJ0}, simply specifies that
\begin{equation}\label{appendix-1dens}
    \int d^3w\tilde{h}_{\mathbf{k}-1}=0.
\end{equation}
Thus, after multiplying \eqref{appendixgyrokineticequation} by $J_0$ and integrating over velocity space, to lowest non-vanishing order we have the equation
\begin{align}\label{StartEquation}
    &\int d^3w \frac{\partial}{\partial t}\left(\tilde{h}_{\mathbf{k}0} - \frac{k_\bot^2w_\bot^2}{2}\tilde{h}_{\mathbf{k}-1} \right) + \int d^3w i\omega_d\left(w_\parallel^2+\frac{w_\perp^2}{2}\right) \tilde{h}_{\mathbf{k}-1} + \int d^3w \PB{\tilde{\varphi}_0}{\overline{h}_0}_\mathbf{k} \nonumber\\ &+ \int d^3w \PB{\overline{\varphi}_0}{\tilde{h}_0}_\mathbf{k} + \int d^3w \PB{\frac{w_\bot^2\nabla^2}{2}\overline{\varphi}_0}{\tilde{h}_{-1}}_\mathbf{k} - \int d^3w \frac{k_\bot^2w_\bot^2}{2}\PB{\overline{\varphi}_0}{\tilde{h}_{-1}}_\mathbf{k} = \nonumber  \\ &\int d^3w \left[ \frac{\partial}{\partial t} + i \omega_* - \frac{i \omega_* \eta k_\bot^2}{2} w_\bot^2 \left( w^2-\frac{3}{2} \right) \right] \tilde{\varphi}_{\mathbf{k}0} f_0 + \int d^3w i \omega_* \eta \left( w^2-\frac{3}{2} \right) \tilde{\varphi}_{\mathbf{k}1} f_0.
\end{align}
The $\tilde{\varphi}_{\mathbf{k}1}$-term vanishes upon integration over velocity space, and so after performing this integration and employing the second lowest order nonzonal quasineutrality condition (remembering the definition \eqref{alphaoperator} for $\hat{\alpha}$)
\begin{equation}\label{appendixquasineutralitynonzonal}
    \int d^3w \tilde{h}_{\mathbf{k}0} - \int d^3w \frac{k_\bot^2w_\bot^2}{2} \tilde{h}_{\mathbf{k}-1} = n(1+\tau)\tilde{\varphi}_{\mathbf{k}0},
\end{equation}
as well as the lowest order zonal quasineutrality condition
\begin{equation}\label{appendixquasineutralityzonal}
    \int d^3w \overline{h}_{\mathbf{k}0} = n\overline{\varphi}_{\mathbf{k}0},
\end{equation}
Equation \eqref{StartEquation} becomes
\begin{align}\label{appendixPhi0}
    &\tau \frac{\partial \tilde{\varphi}_{\mathbf{k}0}}{\partial t} + i \omega_d \left( 2 \tilde{T}_\parallel + \tilde{T}_\perp \right)
    + \PB{\overline{\varphi}_0}{\tau\tilde{\varphi}_0}_\mathbf{k} - \PB{\overline{\varphi}_0}{\nabla_\bot^2\tilde{T}_\perp}_\mathbf{k} \nonumber + \PB{\nabla_\bot^2\overline{\varphi}_0}{\tilde{T}_\perp}_\mathbf{k} - k_\bot^2 \PB{\overline{\varphi}_0}{\tilde{T}_\perp}_\mathbf{k} \\& = i \omega_* \left( 1 - i \eta k_\bot^2 \right) \tilde{\varphi}_{\mathbf{k}0},
\end{align}
where we have cancelled terms, before once again using \eqref{Tnotation}.

Repeating the same process of multiplying \eqref{appendixgyrokineticequation} by $J_0$ and integrating over velocity for the zonal components, at lowest order we find by using \eqref{appendixquasineutralityzonal} the trivial condition that $\overline{\varphi}_{\mathbf{k}0}=\overline{\varphi}_{\mathbf{k}0}$ by virtue of \eqref{appendix-1dens}. Thus we must proceed one order higher in order to arrive at the equation of motion for $\overline{\varphi}_{\mathbf{k}0}$. Then we find the equation
\begin{align}\label{zonalincompleteequation}
    &\int d^3w \frac{\partial}{\partial t} \overline{h}_{\mathbf{k}1} - \int d^3w \frac{k_x^2w_\bot^2}{2} \overline{h}_{\mathbf{k}0} + \int d^3w \PB{\tilde{\varphi}_0}{\tilde{h}_0}_\mathbf{k} + \int d^3w \PB{\frac{w_\bot^2\nabla^2}{2}\tilde{\varphi}_0}{\tilde{h}_{-1}}_\mathbf{k} \\ &- \int d^3w \frac{k_x^2w_\bot^2}{2}\PB{\tilde{\varphi}_0}{\tilde{h}_{-1}}_\mathbf{k} \nonumber  = \int d^3w \frac{\partial \overline{\varphi}_{\mathbf{k}1}}{\partial t} f_0 - \int d^3w \frac{\partial k_x^2\overline{\varphi}_{\mathbf{k}0}}{\partial t} f_0.
\end{align}

Since the first term on both sides of \eqref{zonalincompleteequation} cancel after using the second lowest order zonal quasineutrality condition 
\begin{equation}
    \int d^3w \overline{h}_{\mathbf{k}1} - \int d^3w \frac{k_\bot^2w_\bot^2}{2} \overline{h}_{\mathbf{k}0} = n\overline{\varphi}_{\mathbf{k}1},
\end{equation}
it can be transformed into
\begin{equation}\label{appendixE2}
    - \PB{\tilde{\varphi}_0}{\nabla_\bot^2 \tilde{T}_\bot}_\mathbf{k} + \PB{\nabla_\bot^2 \tilde{\varphi}_0}{\tilde{T}_\perp}_\mathbf{k} - k_x^2 \PB{\tilde{\varphi}_0}{\tilde{T}_\perp}_\mathbf{k} = - \frac{\partial k_x^2 \overline{\varphi}_{\mathbf{k}0}}{\partial t}
\end{equation}
after using \eqref{Tnotation} and \eqref{appendixquasineutralitynonzonal} once more. Now as another consistency check, upon comparing the driving terms in \eqref{appendixPhi0} and \eqref{appendixE2}, which were derived separately, it is indeed found that these differ by an order $\delta^{1/2}$, and so $\overline{\varphi}_{\mathbf{k}0}$ and $\tilde{\varphi}_{\mathbf{k}0}$ are indeed consistent with the ordering separation \eqref{appendixzonal/nonzonal}.

At this point we are essentially done. Equations \eqref{appendixTperp}, \eqref{appendixTpar}, \eqref{appendixPhi0}, and \eqref{appendixE2} constitute a consistent, closed system. Redefining $\omega_{d}\rightarrow\omega_d/4$, dropping the 0-subscripts, before finally including some additional drift wave damping $D_\mathbf{k}$ in accordance with Section \ref{TheDampingOperator}, we have at hand exactly that system \eqref{phi_drift}-\eqref{t_par} which we set out to prove.

\bibliographystyle{jpp}
\bibliography{library.bib}

\end{document}